\documentclass[12pt,preprintnumbers,nofootinbib,letterpaper]{article}
\pdfoutput=1
\usepackage{amsmath,amssymb,amsfonts,cite,color,epsf,epsfig,epstopdf,graphics,graphicx,mathtools,subcaption,physics,verbatim}
\def\thefootnote{*\arabic{footnote}}
\definecolor{ultramarine}{rgb}{0.07, 0.04, 0.56}
\definecolor{cadmiumgreen}{rgb}{0.0, 0.42, 0.24}
\definecolor{indigo(dye)}{rgb}{0.0, 0.25, 0.42}
\usepackage[linktocpage=true,breaklinks]{hyperref}
\hypersetup{
	colorlinks=true,
	citecolor=ultramarine,
	linkcolor=cadmiumgreen,
	urlcolor=indigo(dye),
}

\newcommand{\Mp}{M_{\rm Pl}}
\newcommand{\cs}{c_{\rm s}}

\newcommand{\csr}{c_{{\rm s},sr}}
\newcommand{\cst}{c_{{\rm s},st}}
\newcommand{\cc}{c_{{\rm s},{\rm c}}}
\newcommand{\tc}{\tau_{\rm c}}
\newcommand{\tcmb}{\tau_{\ast}}

\begin{document}

\begin{flushright} {\footnotesize YITP-21-122 \\ IPMU21-0066}  \end{flushright}
\vspace{0.5cm}

\begin{center}

\def\thefootnote{\fnsymbol{footnote}}

{\Large\bf Inflation with \boldmath $0 \leq \cs \leq 1$}
\\[0.7cm]

{Mohammad Ali Gorji$^{1}$, 
Hayato Motohashi$^{2}$,
Shinji Mukohyama$^{1,3}$}
\\[.7cm]

{\small \textit{$^1$Center for Gravitational Physics, Yukawa Institute for Theoretical Physics, Kyoto University, \\ Kyoto 606-8502, Japan
}}\\

{\small \textit{$^2$Division of Liberal Arts, Kogakuin University, 2665-1 Nakano-machi, Hachioji, \\ Tokyo 192-0015, Japan
}}\\

{\small \textit{$^3$Kavli Institute for the Physics and Mathematics of the Universe (WPI), The University of Tokyo, \\ Chiba 277-8583, Japan}}\\

\end{center}

\vspace{.8cm}

\hrule \vspace{0.3cm}


\begin{abstract}
We investigate a novel single field inflationary scenario which allows a transition between a slow-roll k-inflation with $\cs$ of order unity and a ghost inflation with $\cs \simeq 0$, where $\cs$ is the sound speed for the curvature perturbations. 
We unify the two phases smoothly by appropriately taking into account a higher derivative scordatura term, which is always there from the effective field theory point of view but which becomes important only in the $\cs\simeq 0$ regime. 
The model achieves the whole range of $0 \leq \cs \leq 1$ avoiding strong coupling and gradient instability, and allows us to access the $\cs \simeq 0$ regime in a self-consistent manner. We also discuss implications to the formation of primordial black holes.
\end{abstract}
\vspace{0.5cm} 

\hrule
\def\thefootnote{\arabic{footnote}}
\setcounter{footnote}{0}


\newpage
\section{Introduction}

Inflation~\cite{Starobinsky:1980te,Sato:1980yn,Guth:1980zm,Linde:1981mu,Albrecht:1982wi} is an integral part of the modern cosmology which not only solves the problems of the big bang theory such as the horizon, flatness, and monopole problems, but also provides a plausible mechanism to generate seeds for the observed large scale structures in the universe (see \cite{Sato:2015dga} for a review on the historical developments). Apart from these universal features of inflation, there are many different inflationary models~\cite{Martin:2013tda}. In the absence of a full UV complete quantum theory of gravity, observations may decide which model is preferred. 

The CMB observations are in favor of the single field models~\cite{Planck:2018jri} and the most general minimally coupled single field model without higher derivative terms is the so-called slow-roll k-inflation~\cite{ArmendarizPicon:1999rj,Garriga:1999vw,Chen:2006nt}. The power spectrum for the curvature perturbations in k-inflation is given by
\begin{equation}\label{int-PS-cs}
\Delta^2_{\zeta}= \frac{H_{\ast}^2}{8\pi^2\Mp^2} \frac{1}{\epsilon\cs} \,,
\end{equation}
where $H_{\ast}$ is the Hubble expansion rate at the time of horizon crossing and $\Mp=(8\pi{G})^{-1/2}$ is the reduced Planck mass\footnote{We work in unit $c=1=\hbar$ where $c$ is the speed of light in vacuum and $\hbar$ is the reduced Planck constant.}. The parameters $\epsilon$ and $\cs$ are the slow-roll parameter and the sound speed of the curvature perturbations, respectively. 
While the amplitude of the power spectrum on the CMB scales is observationally constrained to the order of $10^{-9}$~\cite{Planck:2018jri}, it can be larger on sub-CMB scales.
A large peak of the power spectrum on small scales leads to the formation of primordial black holes (PBHs)~\cite{1966AZh....43..758Z,Hawking:1971ei}, which can be a candidate for dark matter (for a review, see \cite{Carr:2009jm,Carr:2016drx,Sasaki:2018dmp,Carr:2020gox}).
In the canonical inflation where $c_s=1$, the small-scale enhancement of the power spectrum implies the suppression of $\epsilon$ from the observationally preferred value of order of $10^{-2}$ on CMB scales.
Thus, inflationary models for the PBH production as dark matter should accommodate a transition from $\epsilon < 1$ to $\epsilon \ll 1$ within $\sim 50$ e-folds, leading to an ${\cal O}(1)$ violation of slow-roll~\cite{Motohashi:2017kbs,Passaglia:2018ixg}. 

In the k-inflation, an alternative possibility is to consider small values of the sound speed $\cs\ll1$.
Namely, if $\epsilon$ is kept fixed then the PBH production in the k-inflation requires a transition from $\cs < 1$ to $\cs\ll1$, and such possibility has been discussed recently in various contexts~\cite{Cai:2018tuh,Ballesteros:2018wlw,Kamenshchik:2018sig,Ashoorioon:2019xqc,Ballesteros:2021fsp,Kamenshchik:2021kcw}.
However, a tiny sound speed $\cs\ll1$ requires a special care.
In this limit, the corresponding equilateral bispectrum is characterized by the nonlinear parameter $f_{\rm NL}^{\rm equil} \sim \cs^{-2}$~\cite{Chen:2006nt}. 
This indicates that for very small speed of sound $\cs\ll1$, the setup becomes strongly coupled so that the cubic interactions become comparable or larger than the quadratic ones. In other words, the scale of strong coupling, which scales as a positive power of the sound speed, becomes very small. Although interesting features like large non-Gaussianity and enhancement of power spectrum can be achieved in the regime $\cs\ll1$, the studies in k-inflation are restricted by some critical value of sound speed, below which the linear perturbation theory is no longer applicable~\cite{Cheung:2007st,Motohashi:2019ymr}. 
The lower bound is estimated as $\cs\gtrsim {\cal O}(10^{-2})$~\cite{Ballesteros:2018wlw}.

On the other hand, from the effective field theory (EFT)~\cite{Appelquist:1974tg,Weinberg:1978kz,Georgi:1985kw,Polchinski:1992ed} point of view, higher derivative operators become important in the limit $\cs\ll1$. This point was first noticed in Ref.~\cite{ArkaniHamed:2003uy}, which led to an inflationary scenario known as ghost inflation~\cite{ArkaniHamed:2003uz}. In the ghost inflation scenario, there is an exact or approximate shift symmetry so that $\epsilon\simeq 0$ and $\cs\simeq 0$ while the setup is weakly coupled thanks to the higher derivative operators.
Soft breaking of the shift symmetry was also studied in Ref.~\cite{ArkaniHamed:2003uz,Senatore:2004rj}. 
However, with the exact or approximate shift symmetry, a transition from/to a slow-roll phase is impossible. 
Within the framework of the EFT of inflation~\cite{Creminelli:2006xe,Cheung:2007st} the transition of the sound speed is considered~\cite{Ballesteros:2018wlw,Ashoorioon:2019xqc,Ballesteros:2021fsp}.
Also, a nontrivial evolution of the sound speed is considered phenomenologically~\cite{Cai:2018tuh}.
Yet, no specific model for the transition has been constructed in a self-consistent manner \footnote{We emphasize that both the slow-roll k-inflation and ghost inflation scenarios are self-consistent within their domains of validity. However, a setup which admits transition between them would have a larger domain of validity. For example, the power spectrum in k-inflation, Eq.~\eqref{int-PS-cs}, can be enhanced for small values of $\cs$ but, as we already mentioned, the limit $\cs\to0$ is not accessible in a self-consistent manner. On the other hand, only $\cs\simeq0$ is accessible in the ghost inflation phase while it cannot accommodate $\cs\simeq 1$ regime. Therefore, in a scenario which admits a transition from k-inflation to ghost inflation, it would be possible to achieve a large power spectrum in the allowable limit $\cs\to0$ which was not possible in the pure k-inflation scenario.}. 

A natural question is then whether it is possible to find a specific inflationary model which admits a transition between a slow-roll k-inflation with $\cs$ of order unity and a ghost inflation with $\cs\simeq 0$. This would be an interesting scenario from both theoretical and observational points of view. From theoretical point of view, starting from the slow-roll phase, it allows one to approach not only very small values of the sound speed but also the regime $\cs=0$ in a completely weakly coupled framework. This is a natural extension of the k-inflation with $\cs\ll1$ to the would-be strongly coupled regime. From observational point of view, it enlarges the parameter space that allows one to achieve more interesting features like the formation of PBHs and large non-Gaussianities. 
An implementation of the EFT inspired higher derivative correction into a specific model has been recently explored as the scordatura theory~\cite{Motohashi:2019ymr,Gorji:2020bfl}.
It is thus interesting to apply the scordatura theory to inflation so that it accommodates $0\leq \cs\leq 1$\footnote{In principle, one can further enlarge the parameter region by considering the case with $\cs>1$.  While the standard positivity bound that forbids superluminality assumes the Poincar\'{e} invariant background~\cite{Adams:2006sv} and thus does not directly apply to non-trivial backgrounds, arguments against superluminality still exist even on Lorentz-violating backgrounds~\cite{Dubovsky:2007ac,Shore:2007um,Aoki:2021ffc}. However, these arguments suppose a unitary, Poincar\'{e}-invariant, analytic, and bounded UV completion and thus do not hold if we consider Lorentz-violating UV completion. Also, perturbative superluminality on specific backgrounds does not necessarily indicate pathologies such as acausality (in the sense of the past and the future) with closed timelike curves or ill-posed Cauchy problem~\cite{Babichev:2007dw,Burrage:2011cr,DeRham:2014wnv,Motloch:2015gta}.  Nevertheless, here, we adopt the standard slow-roll phase with $\cs\lesssim 1$ to be conservative.  It is straightforward to generalize our approach in the present paper to construct a model with a transition including the $\cs>1$ phase.}.

In this paper, we construct a simple inflationary model which provides transitions between a slow-roll k-inflation with $\cs$ of order unity and a ghost inflation with $\cs\simeq 0$.  The two phases are smoothly unified by taking into account the higher derivative scordatura term, which arises from the EFT point of view and remedy the strong coupling and gradient instability.  
After setting up our model and notations in \S\ref{sec:model}, we investigate the transition from a slow-roll k-inflation to a ghost inflation and vice versa in unitary gauge in \S\ref{sec:trans}. We then present our explicit single field inflationary model in \S\ref{sec:explicit-model} and study perturbations in \S\ref{sec:pert}. In \S\ref{sec:PBHs}, we address the (im)possibility of the formation of PBHs as dark matter and highlight the role of the scordatura term. We summarize the results in \S\ref{sec:sum}.

\section{The scordatura model}
\label{sec:model}

We consider a simple single field inflationary model defined by the following action
\begin{eqnarray}\label{action-scordatura}
S = \int d^4 x \sqrt{-g} \bigg[ \frac{\Mp^2}{2} R 
+ P(\phi,X) - \frac{\alpha}{2} \frac{(\Box\phi)^2}{M^2} \bigg] \,,
\end{eqnarray}
where $X=\phi_\mu\phi^\mu$, $\Box\phi=\phi^\mu{}_\mu$ with $\phi_\mu = \nabla_\mu\phi$, $\phi_{\mu\nu}=\nabla_\mu\nabla_\nu\phi$. The second term in the action is the k-inflation/essence term from which we can realize the well-known slow-roll k-inflation~\cite{ArmendarizPicon:1999rj,Garriga:1999vw,Chen:2006nt}. The last term, labeled by the constant dimensionless parameter $\alpha$, is the higher derivative term introduced in the context of ghost condensation/inflation~\cite{ArkaniHamed:2003uy,ArkaniHamed:2003uz} and recently named the scordatura term~\cite{Motohashi:2019ymr,Gorji:2020bfl}. In the context of DHOST theories~\cite{Langlois:2015cwa,Crisostomi:2016czh,BenAchour:2016fzp,Takahashi:2017pje,Langlois:2018jdg}, the scordatura term slightly breaks the degeneracy condition~\cite{Motohashi:2014opa,Langlois:2015cwa,Motohashi:2016ftl,Motohashi:2017eya,Motohashi:2018pxg} that prohibits the propagation of the Ostrogradsky ghosts associated with higher derivatives~\cite{Ostrogradsky:1850fid,Woodard:2015zca,Motohashi:2020psc}.\footnote{See also \cite{Aoki:2020gfv} for a generalization of the Ostrogradsky theorem to lower-derivative constrained systems.}  
Unless the degeneracy condition is protected by a fundamental symmetry, it will be eventually broken by higher derivative quantum corrections. Therefore, the scordatura term arises naturally from the EFT point of view. 
In this context, assuming $\alpha={\cal O}(1)$, we regard the mass scale $M$ as the EFT cutoff scale, below which the theory is free from the Ostrogradsky ghost.
Assuming that $M$ is constant and that the EFT is valid all the way down to the present time, the cutoff scale should satisfy $M \lesssim 100 \ {\rm GeV}$~\cite{ArkaniHamed:2005gu}.
If one allows $M$ to vary, on the other hand, this upper bound does not apply to the early universe. We thus assume simply $M/\Mp \ll 1$ throughout the present paper. 
It is worth mentioning that, among all possibilities for the quadratic higher derivative terms like $\phi_{\mu\nu}\phi^{\mu\nu}$ and $\Box\phi \,\phi_\mu \phi^{\mu\nu}\phi_{\nu}$, the term $(\Box\phi)^2$ gives the dominant contribution to the dynamics of perturbations~\cite{Gorji:2020bfl}\footnote{We have also ignored the so-called kinetic braiding term proportional to $\Box\phi$ since it only gives negative contribution to the sound speed squared of the curvature perturbations and, therefore, does not play any significant role in our analysis~\cite{Motohashi:2019ymr}. In the shift symmetric limit, if needed, one can forbid the kinetic braiding term by introducing the $Z_2$ symmetry: $\phi\to -\phi$. }.

The role of the scordatura term was studied in flat and de Sitter backgrounds as a remedy for the (would-be) strong coupling problem of stealth solutions~\cite{Motohashi:2019ymr} and then the study was extended to a more general late-time cosmological background~\cite{Gorji:2020bfl}. For these backgrounds, the stealth solution, which consists of the flat/de Sitter/cosmological background metric and the scalar field with a nontrivial profile~\footnote{See~\cite{Mukohyama:2005rw,Babichev:2013cya,Kobayashi:2014eva,Babichev:2016fbg,Babichev:2017guv,Minamitsuji:2018vuw,BenAchour:2018dap,Motohashi:2019sen,Charmousis:2019vnf,Minamitsuji:2019shy,Minamitsuji:2019tet,Gorji:2019rlm,Takahashi:2020hso,Khoury:2020aya} for stealth black hole solutions with time-dependent scalar profiles.}, is healthy and the EFT describing perturbations around it is weakly coupled all the way up to the cutoff scale $M$~\cite{ArkaniHamed:2003uy}. On the other hand, if we drop the scordatura term by hand then the would-be stealth solution suffers from either infinite strong coupling or gradient instability for scalar field perturbations and thus is not a consistent solution~\cite{Motohashi:2019ymr}. Therefore, the inclusion of the scordatura term is essential for the consistency of the stealth solution and the EFT around it. While the scordatura term appears to introduce an Ostrogradsky ghost, such an apparent ghost is completely benign and does not propagate in the regime of validity of the EFT. The scordatura term is also necessary to make the quasi-static limit well-defined, which implies that the subhorizon observables are inevitably affected by the scordatura \cite{Gorji:2020bfl}. 

While inflationary dynamics in DHOST with/without scordatura has been explored~\cite{Motohashi:2020wxj,Brax:2021qlx,Brax:2021bok}, a transition between slow-roll phase and ghost inflation phase has not been considered.
Also, in the context of ghost inflation, shift symmetric theories were considered in \cite{ArkaniHamed:2003uy,Motohashi:2019ymr,Gorji:2020bfl} and the shift symmetry is broken only softly in \cite{ArkaniHamed:2003uz,Senatore:2004rj}, for which the transition cannot be implemented.
Below, we shall consider a self-consistent inflationary model with the scordatura term to unify the slow-roll and ghost inflation phases.
We consider the possibility of breaking shift symmetry with $P=P(\phi,X)$, which is essential to realize the transition between the two phases as well as to terminate the inflationary regime.

Varying the action~\eqref{action-scordatura} with respect to the metric, we find the Einstein equations
\begin{eqnarray}\label{EEs}
\Mp^2 G^\mu{}_\nu = P \delta^\mu{}_\nu 
- 2 P_{,X} \phi^\mu \phi_\nu + \frac{\alpha}{M^2} \Big[
\Big( \phi^\lambda (\Box\phi)_\lambda
+ \frac{1}{2} (\Box\phi)^2\Big) \delta^\mu{}_\nu 
- \phi^\mu (\Box\phi)_\nu - \phi_\nu (\Box\phi)^\mu
\Big] \,,
\end{eqnarray}
where $G_{\mu\nu} = R_{\mu\nu} - (1/2) R g_{\mu\nu}$ is the Einstein's tensor, and $P_{,X}=\partial P/\partial X$. 
The equation of motion for the scalar field takes the form
\begin{eqnarray}\label{EOM-phi}
2 P_{,X} \Box\phi + 4 \phi^\mu\phi^\nu\phi_{\mu\nu} P_{,XX} 
- P_{,\phi} + 2 X P_{,\phi{X}} + \frac{\alpha}{M^2} \Box^2\phi = 0 \,.
\end{eqnarray}

Following Ref.~\cite{Gorji:2020bfl},
to see the true scaling of the quantities with respect to the scales of physical interest, we define the dimensionless quantities 
\begin{equation}
\phi \equiv \Mp \, \varphi \,, \hspace{1cm} X \equiv M^4 {\rm x} \,, \hspace{1cm} P \equiv M^4 p \,.
\end{equation}
We consider the spatially flat Friedmann-Lema\^{i}tre-Robertson-Walker (FLRW) background
\begin{equation}\label{metric-FRW-BG}
ds^2 = \frac{\Mp^{2}}{M^4} \Big( - d{\tilde t}^2 + a^2 \delta_{ij} d{\tilde x}^i d{\tilde x}^j \Big) \,; \hspace{1cm} (\tilde{t}, \tilde{x}^i) \equiv \frac{M^2}{\Mp} \, (t, x^i) \,,
\end{equation}
where $a({\tilde t})$ denotes the scale factor, ${\tilde t}$ and ${\tilde x}^i$ are the dimensionless cosmic time and dimensionless spatial coordinates respectively which are related to the standard dimensionful time and spatial coordinates $x^\mu=(t,x^i)$ as shown above.

Considering a homogeneous time-dependent VEV $\varphi(t)$ for the dimensionless scalar field, the Friedmann equations take the following form
\begin{eqnarray}\label{Friedmann-b}
3 h^2 = \rho \,, \hspace{1cm}
2 \dot{h} + 3 h^2 = - p \,; \hspace{1cm}
\rho \equiv - p + 2 {\rm x} p_{,{\rm x}} \,,
\end{eqnarray}
where a dot denotes derivative with respect to the dimensionless time $\tilde{t}$, $h=\dot{a}/a$ is the dimensionless Hubble expansion rate, $\rho$ is the dimensionless energy density and ${\rm x}= - \dot{\varphi}^2$.
The equation of motion~\eqref{EOM-phi} for the scalar field becomes
\begin{equation}\label{KG}
\ddot{\varphi} + 3 h \frac{p_{,{\rm x}}}{\rho_{,\rm x}} \dot{\varphi } 
= \frac{\rho_{,\varphi}}{2 \rho_{,\rm x}} \,.
\end{equation}
In obtaining the above equations we neglected effects of the scordatura term since they are suppressed in the limit $M/\Mp\ll 1$ at the background level. Note, however, that they play significant role at the level of perturbations whenever the scalar sound speed is small. 

The line element including the scalar perturbations around the background geometry (\ref{metric-FRW-BG}) is given by (see Appendix~A of Ref.~\cite{Gorji:2020bfl})
\begin{equation}\label{metric-FRW-Perturbations}
ds^2 = \frac{\Mp^2}{M^4} \Big( - ( 1 + 2 A ) d{\tilde t}^2 
+ 2 {\tilde \partial}_i B d{\tilde t} d{\tilde x}^i 
+ a^2 ( 1 + 2 \zeta ) \delta_{ij} d{\tilde x}^i d{\tilde x}^j \Big) \,,
\end{equation}
where $(A,B,\zeta)$ are scalar perturbations, and we have fixed the freedom of spatial diffeomorphism so that the longitudinal part of the spatial metric vanishes. On the other hand, we fix the freedom of temporal diffeomorphism by choosing the unitary gauge in which the scalar field takes the background value as
\begin{equation}\label{scalar-FRW-Perturbations}
 \phi = \Mp \varphi(t)\,,
\end{equation}
Substituting (\ref{metric-FRW-Perturbations}) and (\ref{scalar-FRW-Perturbations}) into (\ref{action-scordatura}), expanding the action up to the quadratic order in perturbations, after integrating out the non-dynamical fields $A$ and $B$, the quadratic action in Fourier space in the limit $M/\Mp\ll 1$ takes the form~\cite{Gorji:2020bfl}
\begin{equation}\label{action}
S^{(2)} \approx \frac{1}{2} \int d{\tilde t} d^3{\tilde k} M^4 
a^3 \, {\cal A} \bigg[ \dot{\zeta}_{\bf k}^2 
- \bigg( \cs^2  \frac{{\tilde k}^2}{a^2} 
+ {\tilde \alpha}^2 \frac{M^2}{\Mp^2}
\frac{{\tilde k}^4}{a^4} \bigg) \zeta_{\bf k}^2
\bigg] \,,
\end{equation}
where we have defined
\begin{eqnarray}\label{cs2-0}
\cs^2 \equiv \frac{p_{,{\rm x}}}{\rho_{,{\rm x}}} 
= \frac{p_{,{\rm x}}}{p_{,{\rm x}}+2 {\rm x} p_{,{\rm x}{\rm x}}} \,,
\hspace{1cm} 
\epsilon \equiv - \frac{\dot{h}}{h^2} \,,
\hspace{1cm} 
{\cal A} \equiv \frac{2\epsilon}{\cs^2} \,, 
\hspace{1cm}
\tilde{\alpha} \equiv \sqrt{\frac{\alpha}{\cal A}} (1-\cs^2) \frac{\dot{\varphi}}{h}\,.
\end{eqnarray}

From the quadratic action~\eqref{action}, the conditions to have no ghost and no gradient instabilities are 
\begin{equation}\label{no-ghost}
{\cal A} > 0 \,, \hspace{1cm} 
\cs^2 \geq 0 \,, \hspace{1cm}
{\tilde\alpha}^2 \geq 0\,.
\end{equation}

\section{Construction in unitary gauge}
\label{sec:trans}

Hereafter, for brevity we call the phase of slow-roll k-inflation a slow-roll phase. Also, we call the phase of ghost inflation a stealth phase. For simplicity we restrict our analysis to a particular functional form for the k-inflation function given by 
\begin{equation}\label{config}
p(\varphi,{\rm x}) = 
{\cal F}_1(\varphi)\, {\rm x} + \frac{1}{2}{\cal F}_2(\varphi)\, {\rm x}^2 - {\rm v}(\varphi)  \,,
\end{equation}
where ${\cal F}_1(\varphi)$ and ${\cal F}_2(\varphi)$ are two arbitrary functions of $\varphi$ and ${\rm v}(\varphi)$ is the potential term. The standard canonical inflation corresponds to the subset ${\cal F}_1(\varphi)=-1/2$, ${\cal F}_2(\varphi)=0$ with the potential ${\rm v}(\varphi)$ satisfying the slow-roll conditions. On the other hand, the so-called ghost inflation corresponds to the shift symmetric subset with constant ${\cal F}_1,{\cal F}_2, {\rm v}$ satisfying ${\cal F}_1>0$ and ${\cal F}_2>0$ so that there is an attractor solution ${\rm x} = - {\cal F}_1/ {\cal F}_2<0$ for the equation $p_{,{\rm x}}=0$ \cite{ArkaniHamed:2003uz}. We are interested in a possible transition between these two inflationary configurations by finding appropriate forms for ${\cal F}_1(\varphi), {\cal F}_2(\varphi), {\rm v}(\varphi)$. In order to do this, we follow the logic used in Ref. \cite{Alberte:2016izw}.

The Friedmann equations~\eqref{Friedmann-b} for the choice \eqref{config} simplify to
\begin{eqnarray}\label{FriedmannEqs-P}
3 h^2 = {\rm x} \mathcal{F}_1 + \frac{3}{2} {\rm x}^2 \mathcal{F}_2 + {\rm v} \,, \hspace{1cm}
\dot{h} = - {\rm x} \big( \mathcal{F}_1 + {\rm x} \mathcal{F}_2 \big) \,,
\end{eqnarray}
from which we find
\begin{eqnarray}\label{Fi-p}
\mathcal{F}_1 = - \frac{ 3 (2-\epsilon) h^2 - 2 {\rm v}}{ {\rm x}} \,, \hspace{1cm}
\mathcal{F}_2 = 2 \frac{ (3-\epsilon) h^2-{\rm v}}{{\rm x}^2} \,.
\end{eqnarray}
Using the above relations, the speed of sound defined in Eq.~\eqref{cs2-0} becomes
\begin{equation}\label{cs2-P}
\cs^2 = \frac{\epsilon h^2}{3 (4-\epsilon) h^2 - 4 {\rm v}} \,.
\end{equation}
Solving it for ${\rm v}$ yields
\begin{equation}\label{v-cs2}
{\rm v} = 3 h^2 - \frac{\epsilon h^2 }{4 \cs^2} (1 + 3 \cs^2) \,.
\end{equation}
Plugging \eqref{v-cs2} into \eqref{Fi-p}, we can express $\mathcal{F}_1,\mathcal{F}_2$ in terms of $c_s^2,\epsilon,h,{\rm x}$.
Considering the unitary gauge 
\begin{equation}\label{unitary-g}
\varphi({\tilde t})= {\tilde t} \,, \hspace{1cm} \dot{\varphi} = 1 \,,
\end{equation}
we find
\begin{eqnarray}\label{F}
\mathcal{F}_1 =
\frac{\epsilon h^2}{2 \cs^2} \big( 1-3 \cs^2 \big) \,, \hspace{1cm}
\mathcal{F}_2 =
\frac{\epsilon h^2}{2 \cs^2} \big( 1-\cs^2 \big) \,.
\end{eqnarray}
After fixing the gauge~\eqref{unitary-g}, 
we see from \eqref{v-cs2} and \eqref{F} that fixing the functional forms of $h,\cs$ is equivalent to fix the functional forms of ${\cal F}_1(\varphi), {\cal F}_2(\varphi), {\rm v}(\varphi)$.
Thus, we only have two free functions.
Taking $\epsilon h^2=-\dot{h}$ and $\cs$ as the two free functions, we can reformulate the dynamics, which is our aim in this subsection. 

First, let us define the slow-roll and stealth regimes in terms of these quantities.
The slow-roll phase is defined by 
\begin{equation}\label{slowroll-phase}
 \epsilon h^2 \neq 0 \,, \hspace{1cm} \cs \neq 0 \,; \hspace{3cm}
\mbox{``slow-roll phase"}.
\end{equation}
(See (\ref{slowroll-phase-p}) for a more precise condition on $\cs$ for the ``slow-roll phase'', based on the behavior of perturbations.) The first condition implies $\epsilon\neq0$ which for $\epsilon\ll1$ guaranties the quasi-de Sitter expansion while the second condition is needed to have a propagating linear scalar perturbations stemming from $P(\phi,X)$ in the action (\ref{action-scordatura}). Moreover, as we see from Eq.~\eqref{F}, we have ${\cal F}_1<0$ for $\epsilon>0$ and $\cs^2>1/3$ which gives the correct sign for the kinetic term linear in ${\rm x}$ in Eq.~\eqref{config}. One may think that it is not possible to achieve $\cs^2<1/3$ during the slow-roll phase as we will have a ghost. However, we note that the effect of $\mathcal{F}_2$ becomes important for the small values of the speed of sound which changes the condition for the absence of ghost. 

The stealth phase is characterized by $\epsilon=0$, $\cs=0$ and $0<{\cal A}<\infty$, where ${\cal A}$ is defined in (\ref{cs2-0}). Therefore, the stealth phase is defined by 
\begin{equation}\label{stealth-phase}
\epsilon = 0 = \cs \,, \hspace{1cm} 
0 < \frac{\epsilon}{\cs^2} < \infty \,; \hspace{3cm}
\mbox{``stealth phase"},
\end{equation}
where the latter condition ensures that ${\cal F}_1$ and ${\cal F}_2$ are non-vanishing and finite in the stealth regime. (See (\ref{stealth-phase-p}) for a more precise condition on $\cs$ for the ``stealth phase'', based on the behavior of perturbations.) For the above configuration, as we see from Eqs.~\eqref{F}, we have ${\cal F}_1>0$ which implies an apparently wrong sign for the linear kinetic term in Eq.~\eqref{config}. This is necessary to have a stealth solution. The apparently wrong sign of the linear kinetic term is of course compensated by the nonlinear kinetic term so that the kinetic term for the curvature perturbations in (\ref{action}) has the correct sign. 

Looking at the slow-roll and stealth configurations, defined by conditions \eqref{slowroll-phase} and \eqref{stealth-phase} respectively, we find that both quantities $\epsilon h^2$ and $\cs$ are non-zero during the slow-roll phase while they vanish during the stealth phase. As we mentioned above, the ratio $\epsilon /\cs^2$ remains finite in the stealth phase when $\epsilon\to 0$ and $\cs\to 0$. One simple consistent choice that guarantees this condition is to impose $\epsilon h^2\propto \cs^2$ during the approach to the stealth phase. On the other hand, we do not need to impose $\epsilon h^2\propto \cs^2$ away from the stealth phase. Considering these points, we thus adopt the following ansatz:
\begin{equation}\label{trans-cond}
\epsilon h^2 = \frac{\beta}{2} \cs^2 \,,
\end{equation}
where $\beta$ is a function of time which should satisfy $\beta>0$ to maintain the condition~\eqref{no-ghost} for the absence of ghost for the scalar perturbations. 
With this parameterization, Eqs.~\eqref{slowroll-phase} and \eqref{stealth-phase} read
\begin{align}\label{slowroll-phase-2}
\cs = \csr \neq 0 \, , \hspace{1cm} 
\beta=\beta_{sr}\,, \hspace{1cm} 
0 < \beta_{sr}<\infty \,&; \hspace{3cm}
\mbox{``slow-roll phase"}, \\
\label{stealth-phase-2}
\cs = 0 \,, \hspace{1cm} 
\beta=\beta_{st}\,, \hspace{1cm} 
0 < \beta_{st} < \infty \,&; \hspace{3cm}
\mbox{``stealth phase"}.
\end{align}

For the above ansatz~\eqref{trans-cond}, Eqs.~\eqref{F} become
\begin{eqnarray}\label{F-sim}
\mathcal{F}_1 =
\frac{\beta}{4} \big( 1-3 \cs^2 \big) \,, \hspace{1cm}
\mathcal{F}_2 =
\frac{\beta}{4} \big( 1-\cs^2 \big) \,.
\end{eqnarray}
In the slow-roll regime \eqref{slowroll-phase-2}, from Eqs.~\eqref{F-sim} we find
\begin{eqnarray}\label{slowroll-config}
\mathcal{F}_{1,sr} =
\frac{\beta_{sr}}{4} \big( 1-3 {c}_{{\rm s},sr}^2 \big) \,, \hspace{1cm}
\mathcal{F}_{2,sr} =
\frac{\beta_{sr}}{4} \big( 1-\csr^2 \big) \,; \hspace{2cm}
\mbox{``slow-roll configuration"}.
\end{eqnarray}
For the standard canonical case with $\csr=1$, we find $\mathcal{F}_{2,sr} =0$ while we have $\csr\neq1$ for the k-inflation. We thus keep the setup general and we do not specify the value of the sound speed during the slow-roll phase $\csr$.
The stealth phase, which is determined by the configuration \eqref{stealth-phase-2}, corresponds to the following conditions
\begin{eqnarray}\label{stealth-config}
\mathcal{F}_{1,st} = \mathcal{F}_{2,st} = \frac{\beta_{st}}{4}  \,; \hspace{4cm}
\mbox{``stealth configuration"}.
\end{eqnarray}

Now, we define the following slow-roll parameters for our inflationary setup
\begin{equation}\label{eta}
\eta \equiv \frac{\dot{\epsilon}}{h \epsilon} = 2 (\epsilon + s) + b \,; \hspace{1cm}
s \equiv \frac{\dot\cs}{h \cs} \,, \hspace{1cm}
b \equiv \frac{\dot{\beta}}{h\beta} \,.
\end{equation}
The parameters $s$ and $b$ characterize time evolution of the sound speed $\cs$ and the parameter $\beta$, respectively. In order to keep the setup under control during the whole period of inflation, we restrict our analysis to the  following slow-roll conditions
\begin{equation}\label{slow-roll-conditions}
(0 \leq) \epsilon \ll 1 \,, \hspace{1cm} |s| \ll 1 \,, \hspace{1cm} |b| \ll 1 \,, \hspace{1cm} |\eta| \ll 1 \,.
\end{equation}
The first condition is necessary to achieve the quasi-de Sitter expansion. Taking this fact into account, only two of the last three conditions should be independently assumed, i.e., assuming $|s| \ll 1$ and $b \ll 1$ automatically implies $\eta\ll{1}$ by definition~\eqref{eta}. There are some other possibilities to have more general models which we do not consider here for the sake of simplicity.

Up to here, we only fixed the values of the slow-roll parameter $\epsilon$ and the sound speed $\cs$ at the slow-roll phase \eqref{slowroll-config} and the stealth phase \eqref{stealth-config}. We did not impose any conditions for the time evolution of these quantities during the transition from a slow-roll phase to a stealth phase or vice versa. In order to do so, we need to specify the functional forms of the sound speed $\cs$ and the parameter $\beta$. Afterward, dynamics of the system will be uniquely determined. For instance, we can find functional forms of the potential and the Hubble expansion rate from Eqs.~\eqref{v-cs2} and \eqref{trans-cond} as follows
\begin{eqnarray}\label{v-h-param}
{\rm v}({\tilde t}) = 3 h^2({\tilde t}) - \frac{1}{8} \beta({\tilde t}) 
\big( 1+ 3\cs^2({\tilde t}) \big) \,, \hspace{1cm}
h({\tilde t}) = - \frac{1}{2} \int d{\tilde t}\, \beta({\tilde t}) \,\cs^2({\tilde t}) \,.
\end{eqnarray}
 
In the following two subsections we shall consider two distinct scenarios: a transition from slow-roll to stealth and vice versa, by specifying appropriate functional forms of the speed of sound $\cs$ and the parameter $\beta$.

\subsection{From slow-roll to stealth}

\begin{figure}[ht] 
	\begin{center}
		\textbf{Transition from slow-roll to stealth}\par\medskip
	\end{center}
	\begin{subfigure}[b]{0.5\linewidth}
		\centering
		\includegraphics[width=0.75\linewidth]{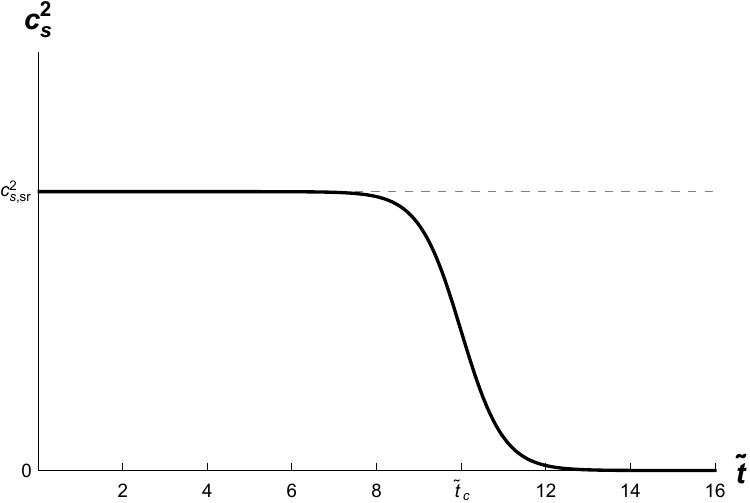}
		\label{fig-cs2t} 
		\vspace{4ex}
	\end{subfigure}
	\begin{subfigure}[b]{0.5\linewidth}
		\centering
		\includegraphics[width=0.75\linewidth]{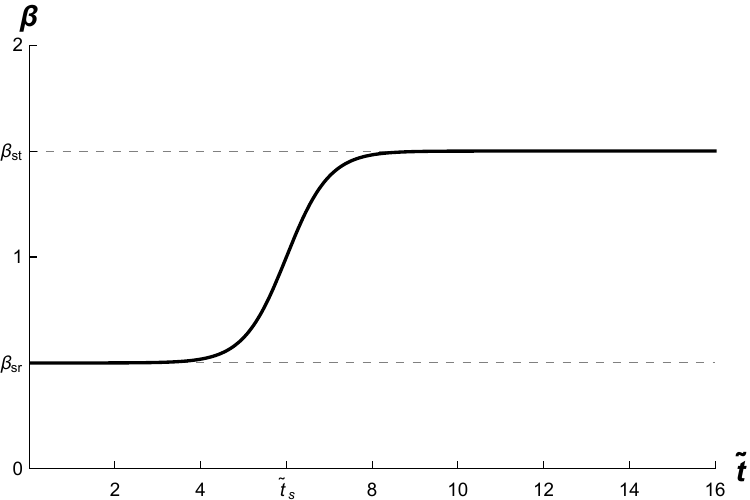} 
		\label{fig-betat} 
		\vspace{4ex}
	\end{subfigure} 
	\begin{subfigure}[b]{0.5\linewidth}
		\centering
		\includegraphics[width=0.75\linewidth]{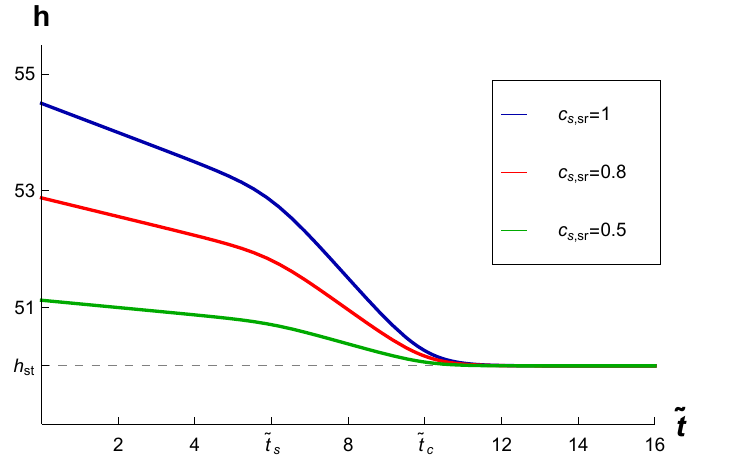} 
		\label{fig-ht} 
	\end{subfigure}
	\begin{subfigure}[b]{0.5\linewidth}
		\centering
		\includegraphics[width=0.75\linewidth]{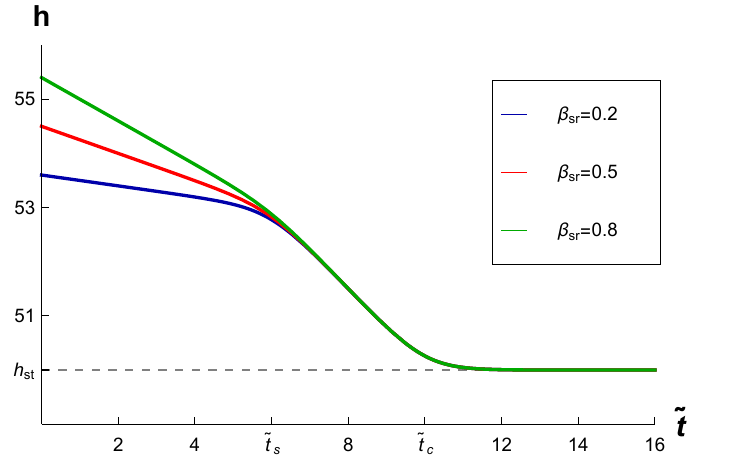} 
		\label{fig-At} 
	\end{subfigure} 
	\caption{The transition from the slow-roll phase to the stealth phase for ${\tilde t}_c=10$, ${\tilde t}_s=6$, $\cst=0$ and $h_{st}=50$. For the first three panels we have set $\beta_{sr}=0.5$ and $\beta_{st}=1.5$ while we have used $\csr=1$ for the bottom right panel. We have the slow-roll phase for $0\leq{\tilde t}\leq4$ and the stealth phase for $12\leq{\tilde t}\leq16$.  The transition takes place in the period $4\leq{\tilde t}\leq12$.}
	\label{Fig:BG-t}
\end{figure}

Let us first consider the case of transition from slow-roll to stealth with $\cs: \csr \to 0$ and $\beta: \beta_{sr} \to \beta_{st}$. As it is clear from Eq.~\eqref{trans-cond}, $\beta_{sr}\ll1$ to ensure small deviation from the exact de Sitter expansion during the slow-roll phase while even $\beta_{st}={\cal O}(1)$ can be achieved during the stealth phase. We thus consider the following ansatze
\begin{align}\label{cs_beta}
\cs^2 &= \cst^2 - (\cst^2 - \csr^2) \cosh({\tilde t}_c)e^{-{\tilde t}} 
\sech({\tilde t}-{\tilde t}_c) \,,
\\ \label{beta}
\beta &= \beta_{st} - (\beta_{st} - \beta_{sr}) \cosh({\tilde t}_s)e^{-{\tilde t}} 
\sech({\tilde t}-{\tilde t}_s) \,, 
\end{align}
where ${\tilde t}_c$ and ${\tilde t}_s$ are constants which determine the time of transitions for $\cs$ and $\beta$ respectively. We have considered non-vanishing value $\cst\ll1$ for the sound speed during the stealth phase to keep our ansatz as general as possible while we will consider $\cst=0$ in all plots and practical purposes. We also consider the case ${\tilde t}_c>{\tilde t}_s$ so that the transition $\beta_{sr} \to \beta_{st}$ takes place before the transition $\csr \to \cst$. The above parameterization is chosen so that $\cs|_{{\tilde t}\ll {\tilde t}_c}=\csr$, as we need for the slow-roll phase with constant sound speed, while $\cs|_{{\tilde t}\gg {\tilde t}_c}=\cst=0$ for the stealth phase. Therefore, the system starts at ${\tilde t}\ll {\tilde t}_s$ from the slow-roll configuration~\eqref{slowroll-config} and approaches to the stealth configuration~\eqref{stealth-config} at ${\tilde t}\gg {\tilde t}_c$ , as shown in Fig.~\ref{Fig:BG-t}. Substituting the ansatze \eqref{cs_beta} and \eqref{beta} into the second equation in \eqref{v-h-param}, we find an explicit expression for the Hubble expansion rate. The constant of integration should be fixed so that $h|_{{\tilde t}\gg {\tilde t}_c}=h_{st}$ where $h_{st}$ is the constant value of the Hubble expansion rate during the stealth phase ${\tilde t}\gg {\tilde t}_c$.

\begin{figure}[ht]
	\begin{center}
		\textbf{Transition from slow-roll to stealth}\par\medskip
	\end{center}
	\begin{subfigure}[b]{0.5\linewidth}
		\centering
		\includegraphics[width=0.75\linewidth]{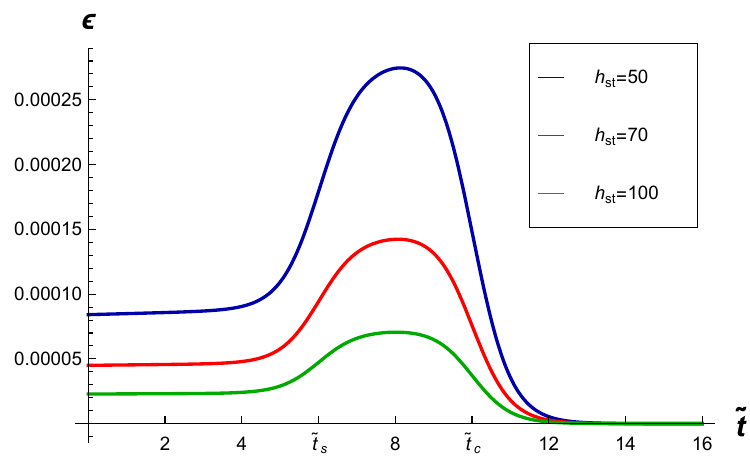} 
		\label{fig-epsilonht} 
		\vspace{4ex}
	\end{subfigure}
	\begin{subfigure}[b]{0.5\linewidth}
		\centering
		\includegraphics[width=0.75\linewidth]{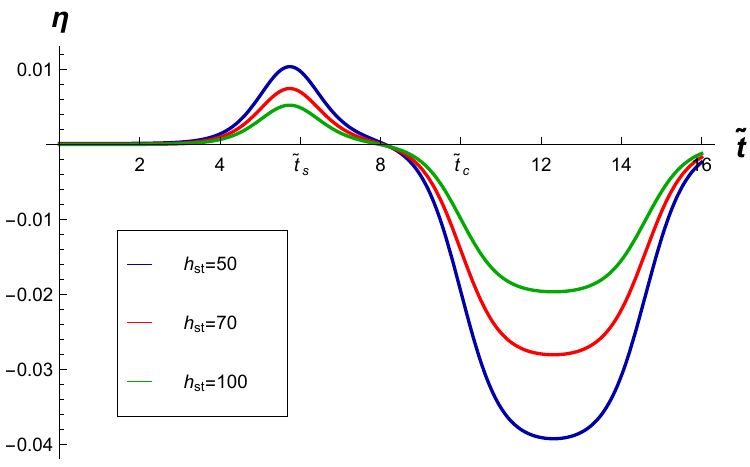}
		\label{fig-etat} 
		\vspace{4ex}
	\end{subfigure}
	\begin{subfigure}[b]{0.5\linewidth}
		\centering
		\includegraphics[width=0.75\linewidth]{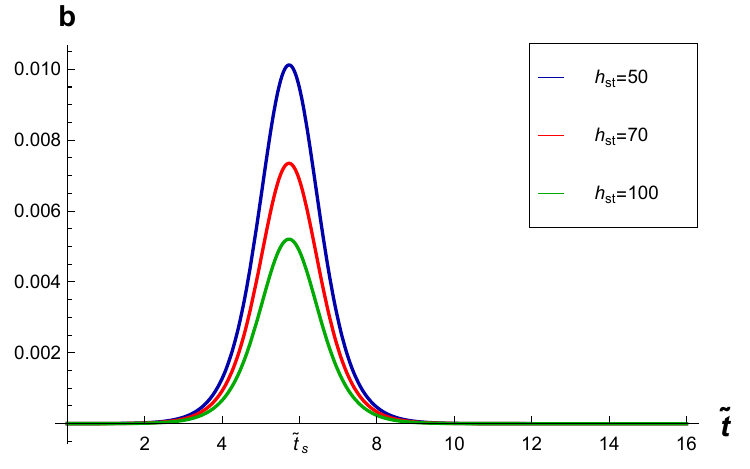}
		\label{fig-epsilonbt} 
		\vspace{4ex}
	\end{subfigure}
	\begin{subfigure}[b]{0.5\linewidth}
		\centering
		\includegraphics[width=0.75\linewidth]{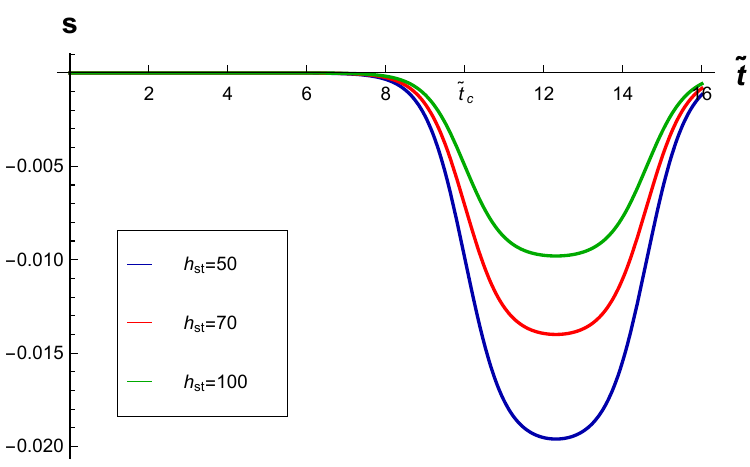} 
		\label{fig-st} 
		\vspace{4ex}
	\end{subfigure} 
	\caption{The slow-roll parameters for the transition from the slow-roll phase to the stealth phase with the parameter set $\beta_{sr}=0.5$, $\beta_{st}=1.5$, $\csr=1$, $\cst=0$, ${\tilde t}_s=6$ and ${\tilde t}_c=10$.}
	\label{Fig:eta-s}
\end{figure}

Having an explicit solution for $h$ in hand, we can obtain all the slow-roll parameters $\epsilon$, $s$, $b$, and $\eta$ as a function of time, which are shown in Fig.~\ref{Fig:eta-s}. We then find $\epsilon:\epsilon_{sr}\to0$, $s:s_{sr}\to0$, $b:b_{sr}\to0$, and $\eta:\eta_{sr}\to0$ with $\epsilon_{sr}\neq0$, $s_{sr}\neq0$, $b_{sr}\neq0$, and $\eta_{sr}\neq0$. From Eq.~\eqref{trans-cond} we see that the value of the slow-roll parameter $\epsilon$ depends on five parameters $\beta_{sr}$, $\beta_{st}$, $\csr$, $\cst$ and $h_{st}$. Fig.~\ref{Fig:eta-s} shows how $\epsilon$ changes by varying $h_{st}$. We emphasize that although we have considered special ansatze~\eqref{cs_beta} and \eqref{beta} for the shape of the transition, the results at both the slow-roll and the stealth phases are quite insensitive to the detailed functional forms of the ansatz. One may consider other appropriate possibilities and the only difference would be in the way that transition takes place.

Note that $h_{st}$ cannot be arbitrarily large as the EFT would break down for a too large value of $h_{st}$. For instance, in the case where we only have the stealth regime, our setup reduces to the ghost inflation scenario \cite{ArkaniHamed:2003uz}. In this case, there is a late time upper bound on the EFT cutoff as $M\lesssim100$ GeV if we assume that the inflaton continues to run with the same velocity even after the reheating. From Eq.~\eqref{metric-FRW-BG} we have $H/M=(M/\Mp)h$, where $H$ is the dimensionful Hubble expansion rate. The COBE normalization on the power spectrum of the curvature perturbations in the ghost inflation implies $H/M\sim{\cal O}(10^{-4})$ \cite{ArkaniHamed:2003uz}. Thus, we find $M={\cal O}(10^{12})$ GeV for $h=100$ which is much greater than $M\lesssim100$ GeV. On the other hand, if we suppose that the cutoff $M$ is time-dependent then we can relax the late time bound $M\lesssim100$ GeV and achieve the value $h=100$.

\subsection{From stealth to slow-roll}
\begin{figure}[ht] 
	\begin{center}
		\textbf{Transition from stealth to slow-roll}\par\medskip
	\end{center}
	\begin{subfigure}[b]{0.5\linewidth}
		\centering
		\includegraphics[width=0.75\linewidth]{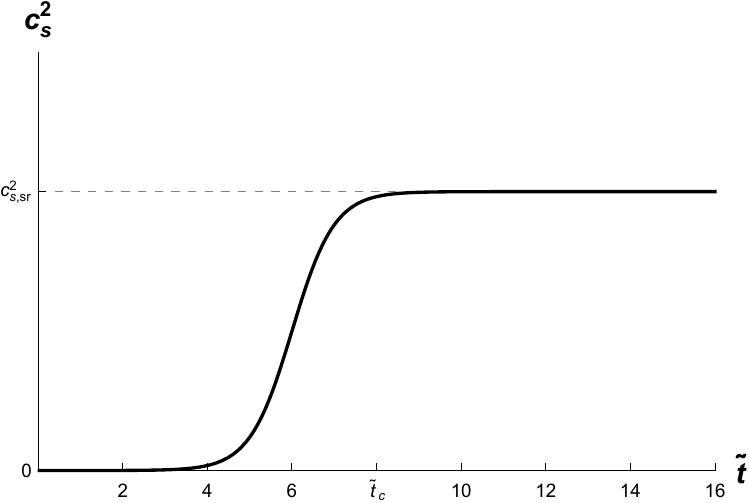}
		\label{fig-cs2t-2} 
		\vspace{4ex}
	\end{subfigure}
	\begin{subfigure}[b]{0.5\linewidth}
		\centering
		\includegraphics[width=0.75\linewidth]{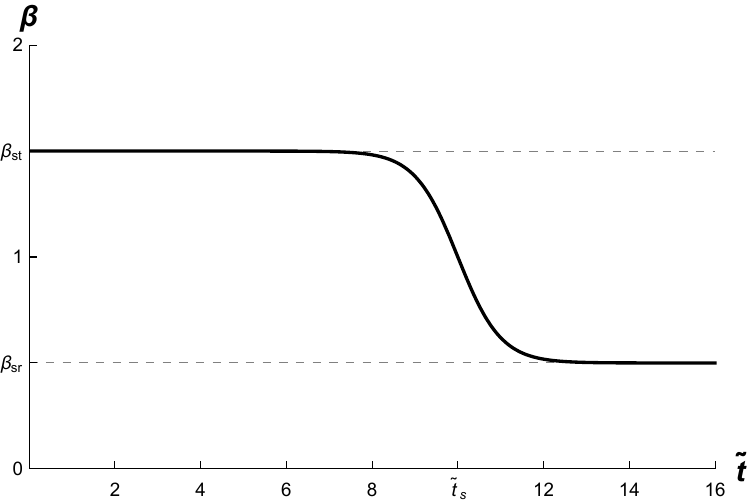} 
		\label{fig-betat-2} 
		\vspace{4ex}
	\end{subfigure} 
	\begin{subfigure}[b]{0.5\linewidth}
		\centering
		\includegraphics[width=0.75\linewidth]{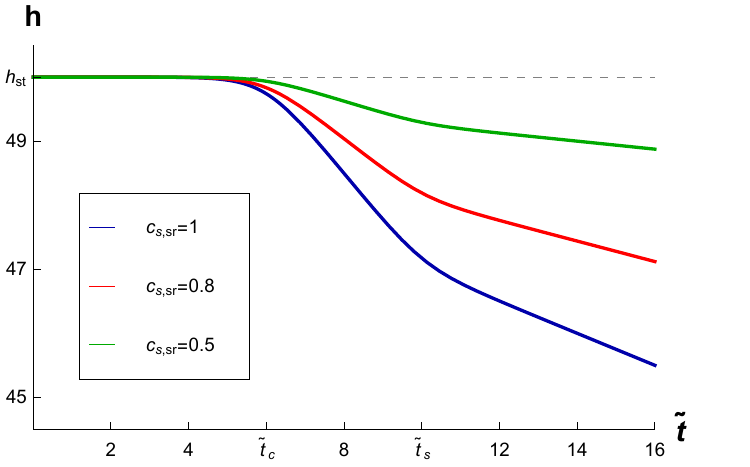} 
		\label{fig-ht-2} 
	\end{subfigure}
	\begin{subfigure}[b]{0.5\linewidth}
		\centering
		\includegraphics[width=0.75\linewidth]{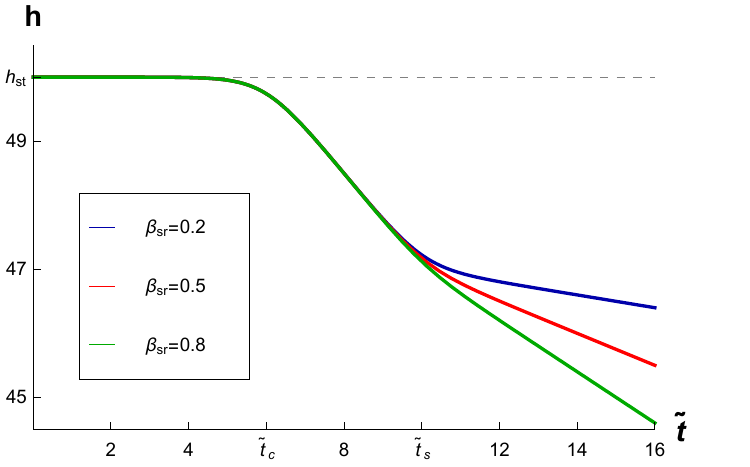} 
		\label{fig-At-2} 
	\end{subfigure} 
	\caption{The transition from the stealth phase to the slow-roll phase with the parameter set ${\tilde t}_c=6$ and ${\tilde t}_s=10$. For the bottom left panel we have set $\beta_{sr}=0.5$, $\beta_{st}=1.5$, $\cst=0$ and $h_{st}=50$ while we have used $c_{sr}=1$, $\beta_{st}=1.5$, $\cst=0$ and $h_{st}=50$ for the bottom right panel. We have the stealth phase for $0\leq{\tilde t}\leq4$ and the slow-roll phase for $12\leq{\tilde t}\leq16$.  The transition takes place in the period $4\leq{\tilde t}\leq12$.}
	\label{Fig:BG-t-2}
\end{figure}
\begin{figure}[ht]
	\begin{center}
		\textbf{Transition from stealth to slow-roll}\par\medskip
	\end{center}
	\begin{subfigure}[b]{0.5\linewidth}
		\centering
		\includegraphics[width=0.75\linewidth]{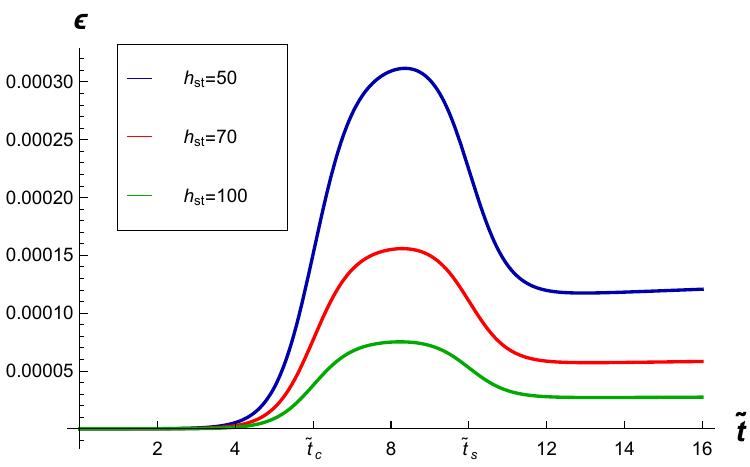} 
		\label{fig-epsilonht-2} 
		\vspace{4ex}
	\end{subfigure}
	\begin{subfigure}[b]{0.5\linewidth}
		\centering
		\includegraphics[width=0.75\linewidth]{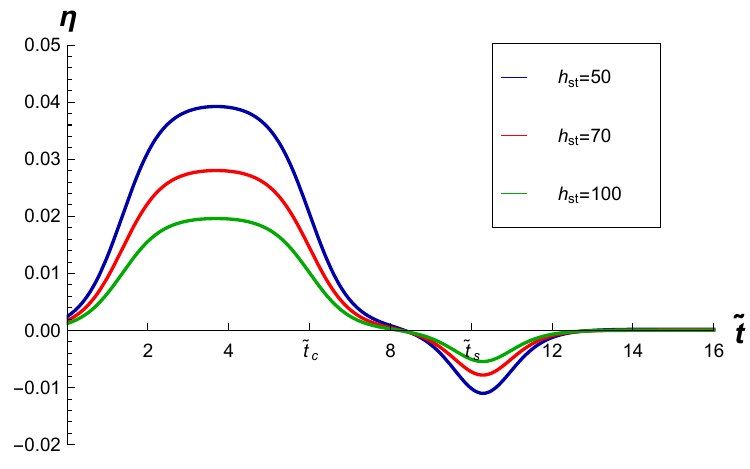}
		\label{fig-etat-2} 
		\vspace{4ex}
	\end{subfigure}
	\begin{subfigure}[b]{0.5\linewidth}
		\centering
		\includegraphics[width=0.75\linewidth]{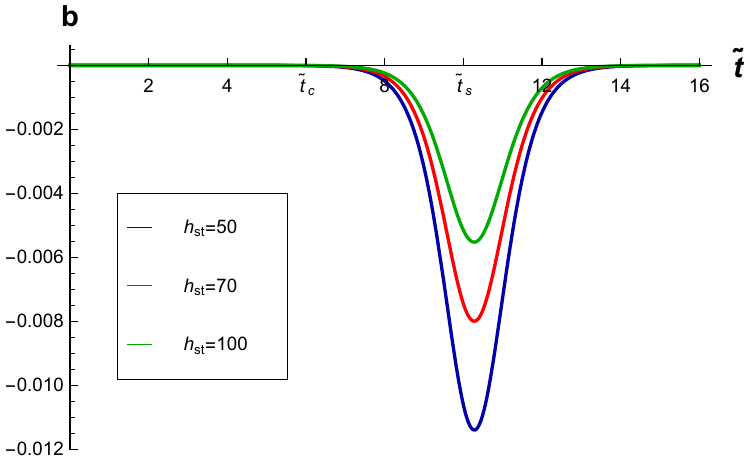}
		\label{fig-epsilonbt-2} 
		\vspace{4ex}
	\end{subfigure}
	\begin{subfigure}[b]{0.5\linewidth}
		\centering
		\includegraphics[width=0.75\linewidth]{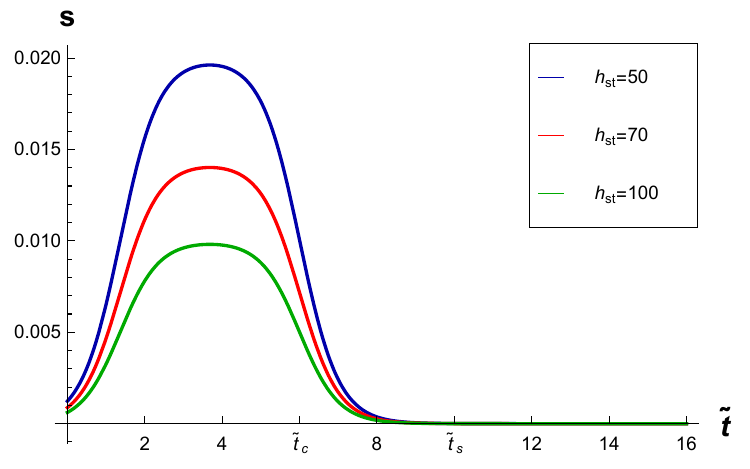} 
		\label{fig-st-2} 
		\vspace{4ex}
	\end{subfigure} 
	\caption{The slow-roll parameters for the transition from the stealth phase to the slow-roll phase with the parameter set $\beta_{sr}=0.5$, $\beta_{st}=1.5$, $\csr=1$, $\cst=0$, ${\tilde t}_c=6$ and ${\tilde t}_s=10$.}
	\label{Fig:eta-s-2}
\end{figure}

Now, we study the case of a transition from stealth to slow-roll with $\cs: \cst \to \csr$ and $\beta: \beta_{st} \to \beta_{sr}$. We consider the following ansatze
\begin{align}\label{cs_beta-2}
\cs^2 &= \csr^2 + (\cst^2 - \csr^2) \cosh({\tilde t}_c)e^{-{\tilde t}} 
\sech({\tilde t}-{\tilde t}_c) \,,
\\ \label{beta-2}
\beta &= \beta_{sr} + (\beta_{st} - \beta_{sr}) \cosh({\tilde t}_s) e^{-{\tilde t}} 
\sech({\tilde t}-{\tilde t}_s) \,.
\end{align}
We consider the case ${\tilde t}_c<{\tilde t}_s$ so that the system starts with the stealth configuration~\eqref{stealth-config} at ${\tilde t}\ll {\tilde t}_c$ and then approaches the slow-roll configuration~\eqref{slowroll-config} at ${\tilde t}\gg {\tilde t}_s$. Substituting ansatze~\eqref{cs_beta-2} and \eqref{beta-2} into the second equation in \eqref{v-h-param}, we find an explicit expression for the Hubble expansion rate and we can find explicit forms of all dynamical quantities. We present the results in Figs.~\ref{Fig:BG-t-2} and \ref{Fig:eta-s-2}.

\section{The inflationary model}
\label{sec:explicit-model}

After constructing the model in unitary gauge in the previous section, we present the explicit form of the model by reintroducing the scalar field. This is similar to the so-called St\"{u}ckelberg trick in EFT of inflation \cite{Cheung:2007st} when we perform a broken time diffeomorphism and introducing the corresponding Nambu-Goldstone boson. The value of the scalar field is fixed in unitary gauge as given by Eq.~\eqref{unitary-g}. We thus reintroduce the scalar field as follows:
\begin{eqnarray}\label{Stuckelberg}
{\tilde t} = \varphi = \frac{\phi}{\Mp} \,.
\end{eqnarray}
From Eqs.~\eqref{F-sim} and \eqref{v-h-param} we then find
\begin{align}\label{F-sim-e}
\mathcal{F}_1(\varphi) &=
\frac{\beta(\varphi)}{4} \big[ 1-3 \cs^2(\varphi) \big] \,, \hspace{1cm}
\mathcal{F}_2(\varphi) =
\frac{\beta(\varphi)}{4} \big[ 1-\cs^2(\varphi) \big] \,,
\\ \label{v-e}
{\rm v}(\varphi) &= 3 h^2(\varphi) - \frac{1}{8} \beta(\varphi) 
\big[ 1+ 3\cs^2(\varphi) \big] \,,
\end{align}
where 
\begin{eqnarray}\label{h-e}
h(\varphi) = - \frac{1}{2} \int d\varphi\, \beta(\varphi) \,\cs^2(\varphi) \,.
\end{eqnarray}
On the desired background specified as an input for the reconstruction procedure in the previous section, the quantities $\cs(\varphi) $ and $\beta(\varphi)$ agree respectively with the sound speed and the ratio of $2\epsilon h^2$ to the sound speed squared (see \eqref{trans-cond}). On the other hand, away from the desired background, $\cs(\varphi) $ and $\beta(\varphi)$ do not agree with these physical quantities in general and are simply functions of $\varphi$ that define the model. Having $\cs(\varphi) $ and $\beta(\varphi)$ in hand, we can find the explicit forms of ${\cal F}_{1,2}$ and ${\rm v}$ in terms of $\varphi$ or $\phi$ and, therefore, we will find an explicit functional form for $P(\phi,X)$ in the action \eqref{action-scordatura} through Eq.~\eqref{config} and the relation $P=M^4 p$. As we have considered different functional forms for $\cs$ and $\beta$ for the transitions between the slow-roll and stealth regimes in the unitary gauge in the previous section, we consider them separately in the following two subsections.

\subsection{From slow-roll to stealth}
In this case, we have considered the ansatze \eqref{cs_beta} and \eqref{beta} in unitary gauge. Reintroducing the scalar field following Eq.~\eqref{Stuckelberg}, we find
\begin{align}\label{cs_beta-e}
\cs^2(\varphi) &= \cst^2 - (\cst^2 - \csr^2) \cosh(\varphi_c)e^{-\varphi} 
\sech(\varphi-\varphi_c) \,,
\\ \label{beta-e}
\beta(\varphi) &= \beta_{st} - (\beta_{st} - \beta_{sr}) \cosh(\varphi_s)e^{-\varphi} 
\sech(\varphi-\varphi_s) \,, 
\end{align}
where $\varphi_c>\varphi_s$. Now, substituting \eqref{cs_beta-e} and \eqref{beta-e} in \eqref{h-e} and fixing the integration constant so that $h|_{{\tilde t}\gg \varphi_c}=h_{st}$ where $h_{st}$ is the constant value of the Hubble expansion rate during the stealth phase $\varphi\gg \varphi_c$, we find the Hubble parameter as a function of the scalar field
\begin{align}\label{h-e-1}
h(\varphi) = h_{{st}} 
+ c_1 \, \varphi 
+ c_2 \, \log \left(e^{2 \varphi _s}+e^{2 \varphi }\right)
+ c_3 \, \log \left(e^{2\varphi_c}+e^{2 \varphi }\right) \,,
\end{align}
where we have defined the following constants
\begin{align}
& c_1 \equiv 
- \frac{1}{2} e^{-2 \left(\varphi _c+\varphi _s\right)} 
\left[
c_{{sr}}^2 \left(e^{2\varphi_c}+1\right)-c_{{st}}^2
\right] 
\left[
\beta_{{sr}}\left(e^{2 \varphi_s}+1\right)-\beta _{{st}}
\right] \,,
\nonumber \\
& c_2 \equiv 
\frac{1}{8} \left(e^{2 \varphi _s}+1\right) e^{-\varphi _c-3 \varphi _s} \left(\beta_{{st}}-\beta_{{sr}}\right) \text{csch}\left(\varphi _c-\varphi _s\right)
\left[
c_{{st}}^2 \left(e^{2 \varphi _s}+1\right)-c_{{sr}}^2 \left(e^{2\varphi _c}+1\right)
\right] \,,
\nonumber \\
& c_3 \equiv
\frac{1}{8} \left(e^{2 \varphi _c}+1\right) e^{-3 \varphi _c-\varphi _s}
\left(c_{{st}}^2-c_{{sr}}^2\right) \text{csch}\left(\varphi _c-\varphi_s\right) 
\left[  \beta_{{sr}} \left( e^{2 \varphi_s} + 1 \right)
- \beta _{{st}} \left(e^{2 \varphi _c}+1\right) \right] \,.
\end{align}

\begin{figure}[ht] 
	\begin{center}
		\textbf{Transition from slow-roll to stealth}\par\medskip
	\end{center}
	\begin{subfigure}[b]{0.5\linewidth}
		\centering
		\includegraphics[width=0.75\linewidth]{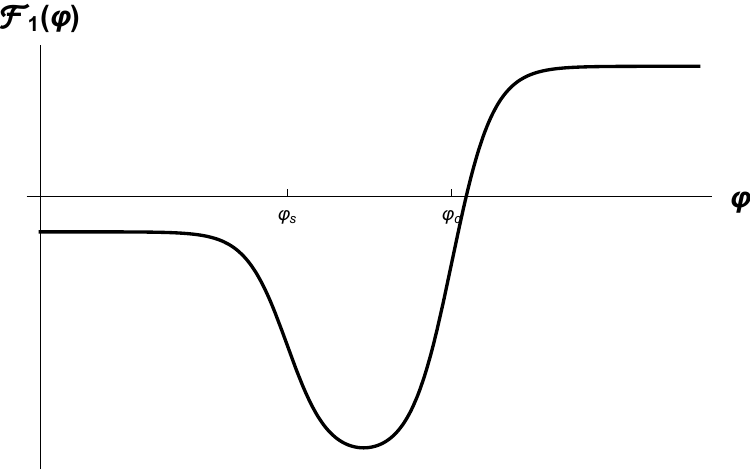}
		\vspace{4ex}
	\end{subfigure}
	\begin{subfigure}[b]{0.5\linewidth}
		\centering
		\includegraphics[width=0.75\linewidth]{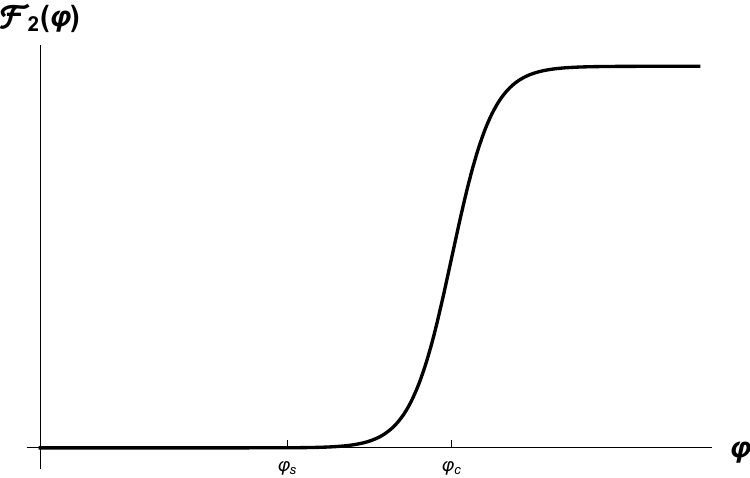} 
		\vspace{4ex}
	\end{subfigure} 
	\begin{subfigure}[b]{0.5\linewidth}
		\centering
		\includegraphics[width=0.75\linewidth]{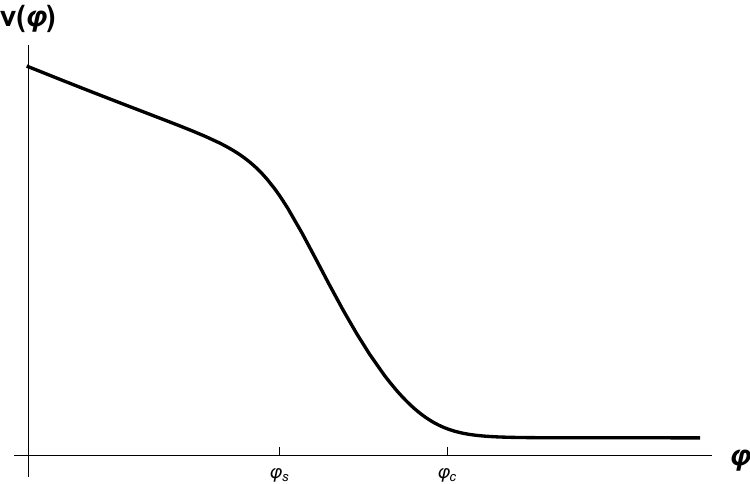} 
	\end{subfigure}
	\begin{subfigure}[b]{0.5\linewidth}
	\centering
	\includegraphics[width=0.6\linewidth]{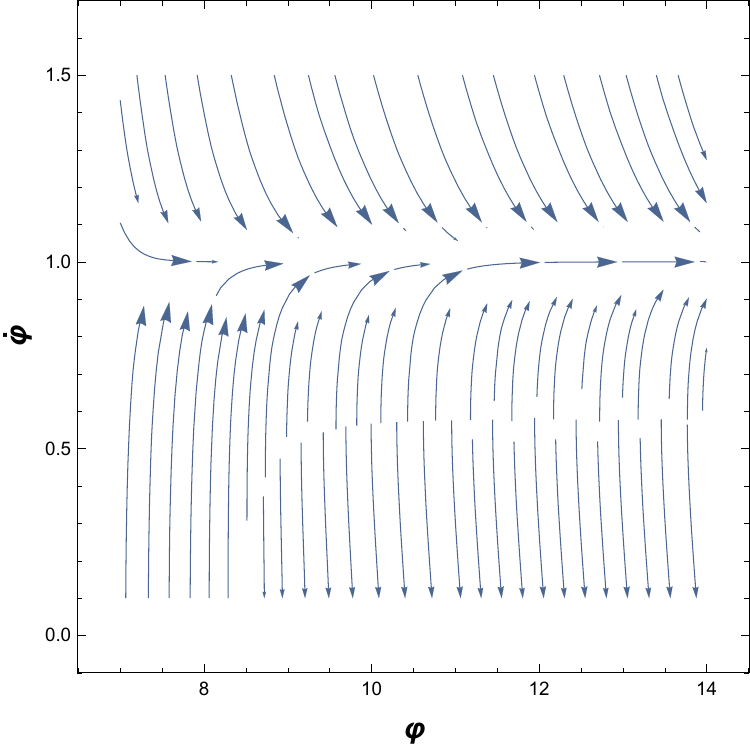}
\end{subfigure}
	\caption{The transition from the slow-roll phase to the stealth phase with the parameter set ${\tilde t}_c=6$, ${\tilde t}_s=10$, $\csr=1$, $\cst=0$, $\beta_{sr}=0.2$, $\beta_{st}=1.5$ and $h_{st}=1$.}
	\label{Fig:F1F2v}
\end{figure}

Having \eqref{cs_beta-e}, \eqref{beta-e}, and \eqref{h-e-1} in hand, we can find ${\cal F}_{1,2}$ and ${\rm v}$ as explicit functions of $\varphi$ or $\phi$ which take complicated forms. The complicated functional forms for ${\cal F}_{1,2}$ and ${\rm v}$ originate from the ansatz \eqref{cs_beta-e} and \eqref{beta-e} that we have considered. One may find other appropriate ansatze which lead to simpler forms. We have plotted ${\cal F}_{1,2}$ and ${\rm v}$ in Figs.~\ref{Fig:F1F2v}. Any analytical functions which have the desired forms as shown in Figs.~\ref{Fig:F1F2v} can be considered as appropriate ansatze.

Moreover, we have plotted the phase portrait in the bottom right panel in Fig.~\ref{Fig:F1F2v}.  For all values of $\dot{\varphi}>1$ independent of the value of $\varphi$, there is always attractor solution while, in the case of $0<\dot{\varphi}<1$, there is only attractor solution at small values of $\varphi$. The reason is that for those large values of $\varphi$, as it can be seen from the potential in Fig.~\ref{Fig:F1F2v}, the setup is already in the ghost inflation phase and $\dot{\varphi}$ cannot be small anymore. These results prove that our setup allows the transition from the slow-roll phase to the stealth phase to be attractor solution and hence provides a viable stable background dynamics for wide range of initial conditions for a reasonably wide range of initial conditions. 

\subsection{From stealth to slow-roll}
For the case of transition from the stealth to slow-roll, reintroducing the scalar field following Eq.~\eqref{Stuckelberg}, from the ansatze \eqref{cs_beta-2} and \eqref{beta-2} we find the following results
\begin{align}\label{cs_beta-e-2}
\cs^2(\varphi) &= \csr^2 + (\cst^2 - \csr^2) \cosh(\varphi_c)e^{-\varphi} 
\sech(\varphi-\varphi_c) \,,
\\ \label{beta-e-2}
\beta(\varphi) &= \beta_{sr} + (\beta_{st} - \beta_{sr}) \cosh(\varphi_s) e^{-\varphi} 
\sech(\varphi-\varphi_s) \,, 
\end{align}
where $\varphi_c<\varphi_s$. Now, substituting \eqref{cs_beta-e-2} and \eqref{beta-e-2} in \eqref{h-e}, we find the Hubble parameter as a function of the scalar field
\begin{align}\label{h-e-2}
h(\varphi) &= h_{st}
+ d_1 \, \varphi 
+ d_2 \, \log \left(\frac{e^{-2\varphi_s}+e^{-2 \varphi }}{e^{-2\varphi_s}+1}\right)
+ d_3 \, \log\left(\frac{e^{-2\varphi_c}+e^{-2 \varphi }}{e^{-2\varphi_c}+1}\right) \,,
\end{align}
where we have defined the following constants
\begin{align}
& d_1 \equiv 
-\frac{1}{2} c_{{sr}}^2 \beta _{{sr}}
 \,,
\nonumber \\
& d_2 \equiv 
\frac{1}{8} \left(e^{2 \varphi_s}+1\right) e^{-\varphi_c-3 \varphi_s} \left(\beta_{{st}}
-\beta_{{sr}}\right) \text{csch}\left(\varphi_c-\varphi_s\right)
\left[
c_{{st}}^2 \left(e^{2 \varphi _c}+1\right)-c_{{sr}}^2 \left(e^{2\varphi_s}+1\right)
\right]
\,,
\nonumber \\
& d_3 \equiv 
\frac{1}{8} \left(e^{2 \varphi_c}+1\right) e^{-3 \varphi_c-\varphi_s}
\left(c_{{st}}^2-c_{{sr}}^2\right) \text{csch}\left(\varphi_c-\varphi_s\right) 
\left[
\beta_{{sr}} \left( e^{2 \varphi_c}+1 \right)
- \beta_{{st}} \left(e^{2 \varphi_s}+1\right)
\right]
\,.
\end{align}

\begin{figure}[ht] 
	\begin{center}
		\textbf{Transition from stealth to slow-roll}\par\medskip
	\end{center}
	\begin{subfigure}[b]{0.5\linewidth}
		\centering
		\includegraphics[width=0.75\linewidth]{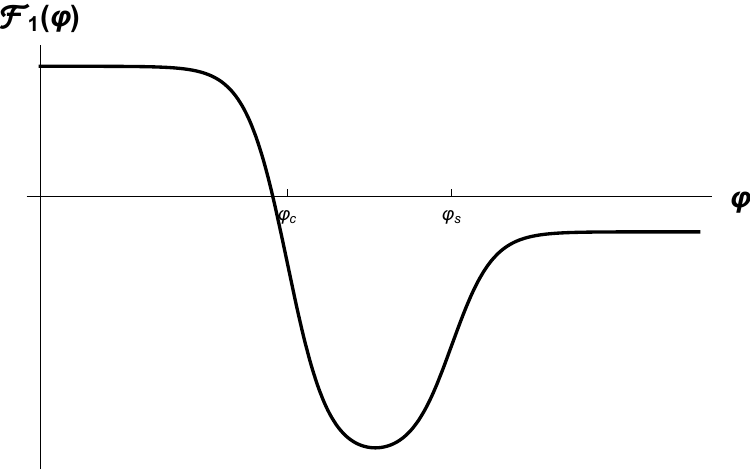}
		\vspace{4ex}
	\end{subfigure}
	\begin{subfigure}[b]{0.5\linewidth}
		\centering
		\includegraphics[width=0.75\linewidth]{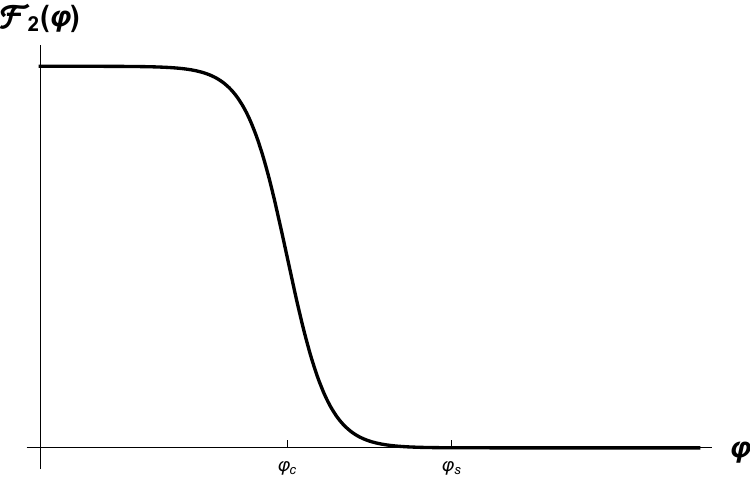} 
		\vspace{4ex}
	\end{subfigure} 
		\begin{subfigure}[b]{0.5\linewidth}
			\centering
			\includegraphics[width=0.75\linewidth]{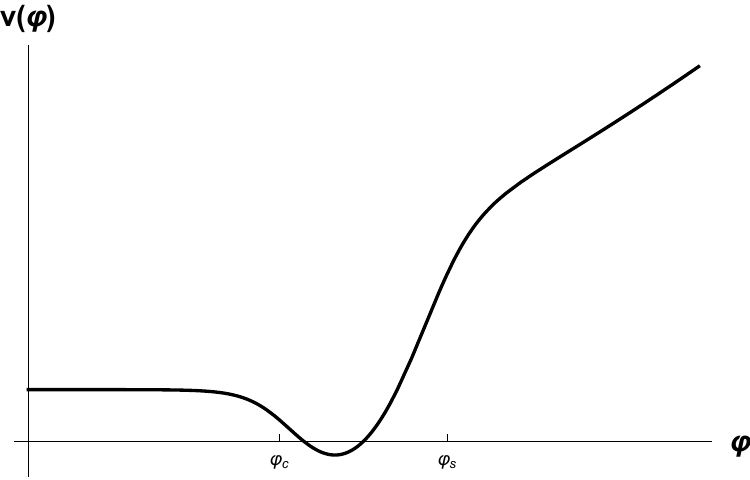} 
		\end{subfigure}
		\begin{subfigure}[b]{0.5\linewidth}
		\centering
		\includegraphics[width=0.6\linewidth]{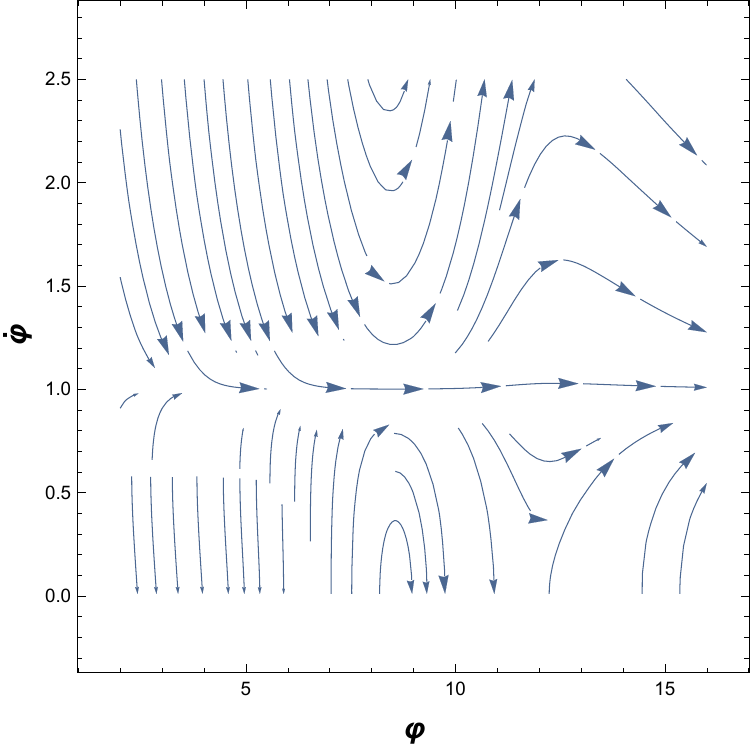}
	\end{subfigure}
	\caption{The transition from the stealth phase to the slow-roll phase with the parameter set ${\tilde t}_c=6$, ${\tilde t}_s=10$, $\csr=1$, $\cst=0$, $\beta_{sr}=0.2$, $\beta_{st}=1.5$ and $h_{st}=1$.}
	\label{Fig:F1F2v-2}
\end{figure}

Having \eqref{cs_beta-e-2}, \eqref{beta-e-2}, and \eqref{h-e-2} in hand, we can find ${\cal F}_{1,2}$ and ${\rm v}$ as explicit functions of $\varphi$ or $\phi$ which are plotted in Figs~\ref{Fig:F1F2v-2}. Any analytical functions which have the desired forms as shown in Figs.~\ref{Fig:F1F2v-2} can be considered as appropriate ansatze.

The phase portrait for transition from the stealth phase to the slow-roll phase is presented in the bottom right panel in Fig.~\ref{Fig:F1F2v-2}. As it can be seen, the transition is an attractor solution for a wide range of initial conditions.

\section{Perturbations}
\label{sec:pert}

To interpret the results for perturbations, let us go back to the original dimensionful coordinates $x^\mu$ defined in Eq.~(\ref{metric-FRW-BG}). The dispersion relation corresponding to the quadratic action Eq.~\eqref{action} takes the form
\begin{equation}\label{DR-total}
\Big(\frac{\omega}{M}\Big)^2 \approx \cs^2 \Big(\frac{k}{a M}\Big)^2
+ {\tilde \alpha}^2 \Big(\frac{k}{aM}\Big)^4 \,.
\end{equation}
In the slow-roll regime with $\cs\neq0$, the scordatura corrections are negligible and we have
\begin{equation}\label{DR-DHOST}
\Big(\frac{\omega}{M}\Big)^2 \approx \csr^2 \Big(\frac{k}{a M}\Big)^2 \,; 
\hspace{3cm} 
\mbox{``slow-roll phase"} .
\end{equation}
Around the stealth solution $\cs\to0$ for $( M/\Mp)\to0$ and the dispersion relation simplifies to
\begin{equation}\label{DR-final}
\Big(\frac{\omega}{M}\Big)^2 \approx {\tilde \alpha}_{st }^2 \Big(\frac{k}{aM}\Big)^4 \,; 
\hspace{3cm}
\mbox{``stealth phase"} ,
\end{equation}
where 
\begin{eqnarray}\label{alpha-tilde-stealth}
{\tilde \alpha}_{st}=\sqrt{\frac{\alpha}{\beta_{st}}}= \mbox{const} \,,
\end{eqnarray}
is the value of the parameter ${\tilde\alpha}$ defined in Eq.~\eqref{cs2-0} around the stealth solution. 

While we can separate the slow-roll and stealth regimes at the background level by means of dynamics of the sound speed and parameter $\beta$, as we discussed in the previous sections, the definitions \eqref{DR-DHOST} and \eqref{DR-final} for the slow-roll and stealth regimes of the perturbations are not very precise. The reason is that for the fixed values of $\cs$ and $\tilde{\alpha}$, due to the dependence on the momenta, we cannot determine which term on the r.h.s.\ of \eqref{DR-total} always dominates. In other words, depending on the momentum, either the first term or second term can dominate. Therefore, at the level of perturbations, we have to define the slow-roll regime~\eqref{DR-DHOST} and stealth regime~\eqref{DR-final} for each mode separately. For the inflationary scenarios, we are mostly interested in the superhorizon modes. Plugging $\omega \sim H$ into the dispersion relation \eqref{DR-total}, we find the horizon crossing mode as
\begin{eqnarray}\label{k-star}
\frac{k_{\ast}}{a_{\ast}M} = 
\frac{c_{{\rm s},\ast}}{\sqrt{2}\tilde{\alpha}_{\ast}} 
\bigg[ \bigg( 1 + 4 \frac{\tilde{\alpha}_{\ast}^2 H_{\ast}^2}{c_{{\rm s},\ast}^4 M^2} \bigg)^{1/2} - 1 \bigg]^{1/2}
\,,
\end{eqnarray}
where all quantities with subscript $\ast$ are evaluated at the horizon crossing conformal time $\tcmb$. Now, we can find the critical value of the sound speed for the horizon crossing mode~\eqref{k-star} at which the two terms on the r.h.s.\ of the dispersion relation \eqref{DR-total} become comparable. Considering that this happens at the conformal time $\tau_{\rm sc}$, we find
\begin{eqnarray}\label{cs-star}
c_{{\rm s},{\rm sc} \ast} = 
\frac{c_{{\rm s},\ast}}{\sqrt{2}} \frac{\tilde{\alpha}_{\rm sc}}{\tilde{\alpha}_{\ast}} \frac{a_{\ast}}{a_{\rm sc}}
\bigg[ \bigg( 1 + 4 \frac{\tilde{\alpha}_{\ast}^2 H_{\ast}^2}{c_{{\rm s},\ast}^4 M^2} \bigg)^{1/2} - 1 \bigg]^{1/2}
\,.
\end{eqnarray}
Thus, we define the slow-roll phase and stealth phase for perturbations as follows:
\begin{align}\label{slowroll-phase-p}
c_{{\rm s},{\rm sc} \ast} \ll \cs \leq 1&; \hspace{3cm}
\mbox{``slow-roll phase"}, \\
\label{stealth-phase-p}
0 \leq \cs \ll c_{{\rm s},{\rm sc}\ast}&; \hspace{3cm}
\mbox{``stealth phase"},
\end{align}
which is consistent with the definitions~\eqref{DR-DHOST} and \eqref{DR-final} when we consider superhorizon modes.
For $\cs \sim c_{{\rm s},{\rm sc}\ast}$, both the k-inflation term and the scordatura term in the action \eqref{action-scordatura} contribute to the dispersion relation of the curvature perturbations. 

\begin{figure}[ht] 
	\begin{center}
		\textbf{Transition from stealth to slow-roll}\par\medskip
	\end{center}
	\begin{subfigure}[b]{0.5\linewidth}
		\centering
		\includegraphics[width=0.9\linewidth]{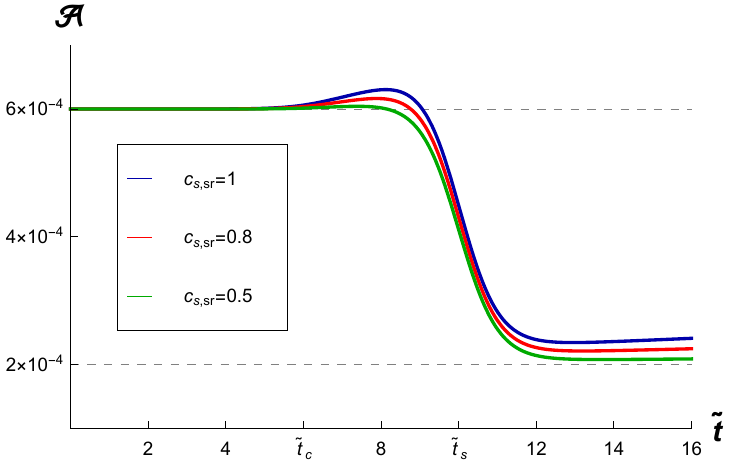}
		\vspace{4ex}
	\end{subfigure}
	\begin{subfigure}[b]{0.5\linewidth}
		\centering
		\includegraphics[width=0.9\linewidth]{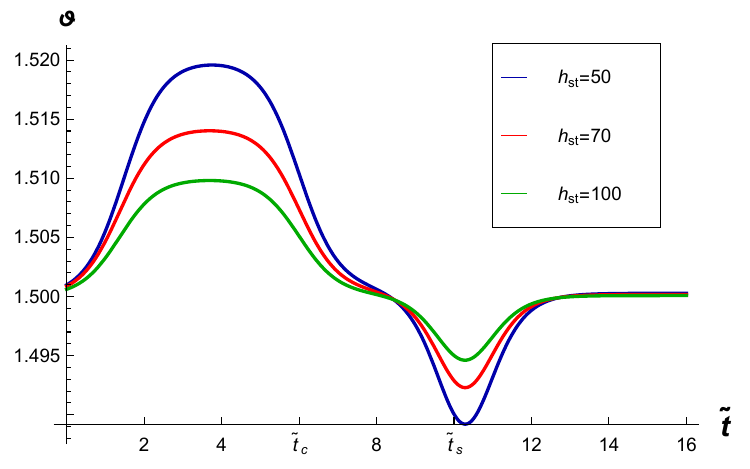} 
		\vspace{4ex}
	\end{subfigure}
	\caption{Left panel: $\beta_{sr}=0.5$, $\beta_{st}=1.5$, ${\tilde t}_c=10$, ${\tilde t}_s=6$ and $h_{st}=50$. Right panel: $\csr=1$, ${\tilde t}_c=6$ and ${\tilde t}_s=10$.}
	\label{Fig:vartheta-2}
\end{figure}

Having defined the slow-roll and stealth regimes for the perturbations, now, our aim is to find the power spectrum for the curvature perturbations ${\zeta}_{\bf k}$. Before finding the power spectrum, let us make a comment on the evolution of the curvature perturbations on superhorizon scales. This is important as superhorizon modes serve as seeds for large scale structures in the universe. Considering the long wavelength limit ${\tilde k}\to 0$ in the action \eqref{action}, we find the following solution
\begin{equation}\label{superhorizon}
{\zeta}_{\bf k} \approx \mbox{const} + \int \frac{d{\tilde t}}{a^3 {\cal A}} \,, \hspace{2cm} 
\mbox{for superhorizon modes}.
\end{equation}
In the case of the standard slow-roll inflation we have ${\cal A}\propto\epsilon$ which is slowly varying during the whole inflationary period and the last term in the r.h.s.\ decays through the exponential expansion. 
Consequently, the curvature perturbation remains constant on superhorizon scales.
However, this is not always the case, even in the canonical inflation.
In ultra-slow-roll inflation~\cite{Tsamis:2003px,Kinney:2005vj} and some subclass of constant-roll inflation~\cite{Martin:2012pe,Motohashi:2014ppa,Motohashi:2019rhu}, $\epsilon$ decays quicker than $a^{-3}$, which implies that the second term is growing outside the horizon.
In the present case, we need to check the evolution of ${\cal A}$ at the both phases and also during the transition. 
From Figs.~\ref{Fig:vartheta} and \ref{Fig:vartheta-2}, we see that time variation of ${\cal A}$ is slow and therefore the second term in the r.h.s.\ of superhorizon solution~\eqref{superhorizon} decays as usual. 
Thus, the curvature perturbation is frozen on superhorizon scales in both cases of the transition between slow-roll and stealth phases.

In order to quantize the system, we rewrite the action \eqref{action} in terms of the conformal time $\tau = \int dt/a(t)$ as
\begin{equation}\label{action-canonical}
S^{(2)} \approx \frac{1}{2M^2}\int d{\tau} d^3k
\bigg[ \bar{\zeta}_{\bf k}'^2 - \bigg( \cs^2{ k}^2
+ \frac{{\tilde \alpha}^2 k^4}{a^2 M^2} - \frac{z''}{z} \bigg) \bar{\zeta}_{\bf k}^2
\bigg] \,,
\end{equation}
where a prime denotes derivative with respect to the conformal time and we have defined the canonical field
\begin{eqnarray}\label{CP-canonical}
\bar{\zeta}_{\bf k} \equiv z\, \zeta_{\bf k}\,; \hspace{1cm}
z \equiv \Mp a \sqrt{\beta} (M/H) \,.
\end{eqnarray}

Promoting the canonically normalized variable $\bar{\zeta}_{\bf k}$ to an operator and expanding it in terms of the creation and annihilation operators as
\begin{eqnarray}\label{h-op}
\bar{\zeta}_{\bf k}(\tau)= \bar{\zeta}_{k}(\tau) a_{\textbf{k}} 
+ \bar{\zeta}_{k}^{*}(\tau) a_{-{\bf k}}^{\dagger} \,; \hspace{1cm}
\big[ a_{\bf k}, a_{{\bf k}'}^{\dagger} \big]= (2\pi)^3 \delta({\bf k}-{\bf k}') \,,
\end{eqnarray}
we find the following equation for the mode function
\begin{eqnarray}\label{EOM-nu}
\bar{\zeta}''_k + \bigg( \cs^2{ k}^2
+ \frac{{\tilde \alpha}^2 H^2}{M^2} k^4 \tau^2 
- \frac{\vartheta^2-1/4}{\tau^2} \bigg) \bar{\zeta}_k = 0 \,;
\hspace{1cm} 
\vartheta \equiv \frac{3}{2} + \epsilon + \eta - s \,,
\end{eqnarray}
where we have expanded the result of $\vartheta$ for the small values of $\epsilon, \eta, s, b$ and we have erased $b$ by using Eq.~\eqref{eta}. The exact form of $\vartheta$ for transition between slow-roll and stealth phases are plotted in Figs.~\ref{Fig:vartheta} and \ref{Fig:vartheta-2}. In obtaining the above result, we have also assumed that $\dot{b}/bH={\cal O}(\epsilon,\eta,b,s)$ and therefore we have neglected the time variation of the parameter $b$. Had we assumed a constant $\beta$, $\vartheta$ would be independent of $\eta$. Thus, we have to assume that $\beta$ depends on time in the ansatz Eq.~\eqref{trans-cond} to be able to recover the slow-roll phase. Assuming constant $\beta$ is another interesting possibility which can be considered separately.

In order to solve Eq.~\eqref{EOM-nu} we need to take into account the time dependence of $\cs$ and $\tilde{\alpha}$. Then, it is not possible to find an analytical solution for Eq.~\eqref{EOM-nu}. However, if the values of $\cs$ and $\tilde{\alpha}$ do not change significantly in time, we can find an analytical solution by treating them approximately constant. In this case, the positive frequency Bunch-Davies solution for Eq.~\eqref{EOM-nu} is given by \cite{Aoki:2020ila}
\begin{equation}\label{Whittaker} 
\bar{\zeta}_k = \Big(\frac{M}{{\tilde \alpha}H}\Big)^{1/2} \, \frac{e^{-\frac{\pi \cs^2 M}{8{\tilde \alpha}H}}}{\sqrt{-2\tau}k} \,
W\left( \frac{i \cs^2 M}{4{\tilde \alpha}H}, \frac{\vartheta}{2}, 
- \frac{i{\tilde \alpha}Hk^2\tau^2}{M} \right) \,,
\end{equation}
where $W(\kappa,\lambda,z)$ is the Whittaker function. This solution can be used for any constant values of $\cs$ and $\tilde{\alpha}$. 

In the limit ${\tilde \alpha}\to0$, the solution~\eqref{Whittaker} reduces to the Hankel function as follows
\begin{align}\label{zeta-slow-roll} 
&\bar{\zeta}_k\big{|}_{{\tilde \alpha}\to 0} = e^{i\theta_1} \frac{\sqrt{-\pi\tau}}{2} 
 {\rm H}^{(1)}_{\vartheta_1}(x)  \,;
&x \equiv -\cs k\tau \,, \hspace{.5cm}
\vartheta_1 \equiv \vartheta\big{|}_{{\tilde \alpha}\to 0} \,,
\end{align}
where the value of $\theta_1$ in the phase factor is irrelevant for our purpose as we are interested in the cosmological correlation functions.

In the limit $\cs\to0$, the solution~\eqref{Whittaker} reduces to the following solution
\begin{align}\label{zeta-stealth} 
&\bar{\zeta}_k\big{|}_{\cs\to 0} 
= e^{i\theta_2} \frac{\sqrt{-\pi\tau}}{2\sqrt{2}} {\rm H}^{(1)}_{\vartheta_2/2} 
\left(y\right) \,; 
&y \equiv \frac{{\tilde \alpha} H k^2\tau^2}{2M} \,, \hspace{.5cm}
\vartheta_2 \equiv \vartheta\big{|}_{\cs\to 0} \,,
\end{align}
in agreement with \cite{ArkaniHamed:2003uz}.

In order to avoid discontinuity, we have to impose junction conditions for $\zeta_{\bf k}$ and its conjugate momentum ${\cal A}\,\dot{\zeta}_{\bf k}$. Considering these junction conditions between the initial phase and final phase at the time $\tau=\tc$, we have
\begin{eqnarray}\label{JC0}
\frac{1}{z_{i}} \bar{\zeta}^{i}_{\bf k} \Big{|}_{\tau=\tc} = 
\frac{1}{z_{f}} \bar{\zeta}^{f}_{\bf k} \Big{|}_{\tau=\tc} \,, 
\hspace{1cm} 
h_i {\cal A}_{i} \Big(\frac{1}{z_{i}} \bar{\zeta}^{i}_{\bf k}\Big)' \Big{|}_{\tau=\tc} = 
h_f {\cal A}_{f} \Big(\frac{1}{z_{f}} \bar{\zeta}^{f}_{\bf k}\Big)' \Big{|}_{\tau=\tc} \,, 
\end{eqnarray}
where
\begin{eqnarray}
h_i {\cal A}_{i} = \frac{\beta_{i}}{h_{i}} \,, \hspace{.5cm} 
z_{i} = \frac{\Mp^2}{M} \frac{\sqrt{\beta_{i}}}{h_i^2} \, \frac{1}{-\tau} \,, 
\hspace{1.5cm} 
h_f {\cal A}_{f} = \frac{\beta_{f}}{h_{f}} \,,  \hspace{.5cm} 
z_{f} = \frac{\Mp^2}{M} \frac{\sqrt{\beta_{f}}}{h_f^2} \, \frac{1}{-\tau} \,, 
\end{eqnarray}
in which we have ignored slow-roll suppressed corrections. We also note that the parameter $\beta$ is considered to have a constant value during the stealth phase and, therefore, we ignore its time evolution during the stealth phase. Moreover, we neglect its time evolution even during the slow-roll phase as it gives a correction proportional to $b$ which is small in our setup. Thus, we find
\begin{eqnarray}\label{JC}
\bigg[ \Big(\frac{\beta_{f}}{\beta_{i}}\Big)^{1/2} 
\Big(\frac{H_{i}}{H_{f}}\Big)^{2}
\, \bar{\zeta}^{i}_{\bf k} -
\bar{\zeta}^{f}_{\bf k} \bigg]_{\tau=\tc} &=& 0 \,, 
\nonumber \\
\bigg[ \bar{\zeta}^{i}_{\bf k}{}' 
- \Big( \frac{\beta_{f}}{\beta_{i}}\Big)^{1/2} \frac{H_f}{H_i}  \, 
\bar{\zeta}^{f}_{\bf k}{}' \bigg]_{\tau=\tc} 
&=&
\bigg[ \bigg( 1 - \frac{\beta_{f}}{\beta_{i}} \frac{H_{i}}{H_{f}} \bigg) 
\frac{\bar{\zeta}^{i}_{\bf k}{}}{-\tau} \bigg]_{\tau=\tc} \,.
\end{eqnarray}
We will use the above junction conditions to study transition between slow-roll and stealth phases in the next two subsections.

\subsection{From slow-roll to stealth}\label{subsec-SRtoST}

Let us first consider the case when inflation starts with an initial slow-roll phase and it transits into a stealth phase.

In the slow-roll regime for $\tau\ll\tc$, the positive frequency Bunch-Davies mode function is given by Eq.~\eqref{zeta-slow-roll} as
\begin{align}\label{zeta-SR-initial} 
&\bar{\zeta}^{sr,i}_k = \frac{\sqrt{-\pi\tau}}{2} 
{\rm H}^{(1)}_{\nu}(x_{sr}) \,; 
&\nu \equiv \vartheta_{1,sr} \,,
\end{align}
where $x_{sr} \equiv -\csr k\tau $.
From Eq.~\eqref{CP-canonical}, we see that the two-point correlation function for the curvature perturbations at the time $\tau\ll\tc$ is given by
\begin{eqnarray}\label{PS-CP-SR}
\langle \zeta_{\bf k}(\tau) \zeta_{{\bf k}'}(\tau) \rangle|_{\tau\ll\tc} = z_{sr}^{-2} 
\big{|} \bar{\zeta}^{sr,i}_k \big{|}^2 (2\pi)^{3} \delta^{(3)}({\bf k}+{\bf k}') \,,
\end{eqnarray}
where $z_{sr}$ denotes the value of $z$ at the slow-roll phase. The dimensionless power spectrum $\Delta^2_{\zeta}(k)$ defined as 
\begin{eqnarray}\label{PS-def}
\langle \zeta_{\bf k}(\tau) \zeta_{{\bf k}'}(\tau) \rangle \equiv (2\pi^2/k^3) \Delta^2_{\zeta}(k)|_{\tau\ll\tc} (2\pi)^3 \delta({\bf k}+{\bf k}') \,,
\end{eqnarray}
is obtained as
\begin{equation}\label{PS-zeta-sr}
\Delta^2_{\zeta}(k)|_{\tau\ll\tc} = \frac{H_{sr}^2}{4\pi^2\Mp^2} \Big(\frac{H_{sr}}{M}\Big)^2
\frac{1}{\beta_{sr}\csr^3} \Big(\frac{\Gamma(\nu)}{\Gamma(3/2)}\Big)^2 
\Big(\frac{x_{sr}}{2}\Big)^{3-2\nu} \, ,
\end{equation}
where we have used the asymptotic behavior of the Hankel function ${\rm H}^{(1)}_{\vartheta}(x)\approx-i (2/x)^\vartheta(\Gamma(\vartheta)/\pi)$ for $x\ll1$. 
The spectral tilt at the slow-roll regime is given by
\begin{equation}\label{sr-tilt}
n_{s}^{sr,i} -1 \equiv \frac{d\ln\Delta^2_{\zeta}|_{\tau\ll\tc}}{d\ln{k}} = 3-2\nu \,,
\end{equation}
which shows that the power spectrum is almost scale-invariant. 

\begin{figure}[ht] 
	\begin{center}
	\textbf{Transition from slow-roll to stealth}\par\medskip
\end{center}
	\begin{subfigure}[b]{0.5\linewidth}
		\centering
		\includegraphics[width=0.9\linewidth]{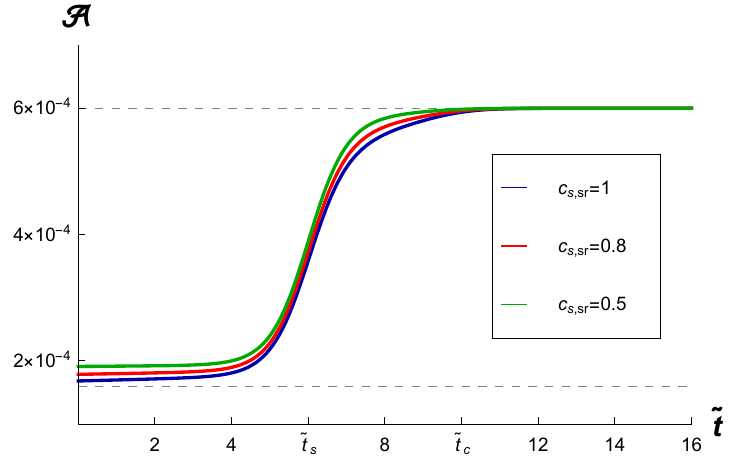}
		\vspace{4ex}
	\end{subfigure}
	\begin{subfigure}[b]{0.5\linewidth}
		\centering
		\includegraphics[width=0.9\linewidth]{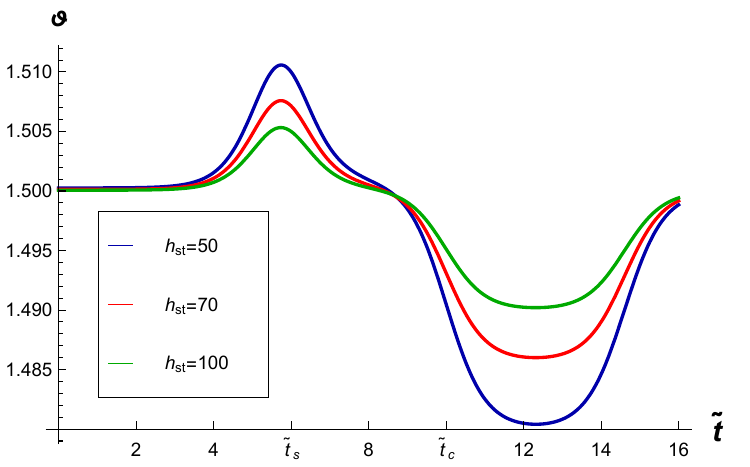} 
		\vspace{4ex}
	\end{subfigure}
	\caption{Left panel: $\beta_{sr}=0.5$, $\beta_{st}=1.5$, ${\tilde t}_c=10$, ${\tilde t}_s=6$ and $h_{st}=50$. Right panel: $\csr=1$, ${\tilde t}_c=6$ and ${\tilde t}_s=10$.}
	\label{Fig:vartheta}
\end{figure}

In the stealth phase the mode function is a general solution of Eq.~\eqref{EOM-nu} with $\cs=0$ which is given by
\begin{align}\label{zeta-ST-final} 
&\bar{\zeta}^{st,f}_k = 
\frac{\sqrt{- \pi \tau }}{2} \left[ 
c^{st}_2 {\rm H}_{\mu/2}^{(1)}\left(y_{st}\right)
+c^{st}_1 {\rm H}_{\mu/2}^{(2)}\left(y_{st}\right)\right] \,;
& \mu \equiv \vartheta_{2,st} \,,
\end{align}
where coefficients $c^{st}_\ell$ with $\ell=1,2$ are some constants. The parameter $\vartheta:\nu \to \mu$ is plotted in the right panel of Fig.~\ref{Fig:vartheta}. It is worth mentioning that since $y_{st} = ({\tilde \alpha}_{st} H_{st}/2M)k^2\tau^2$, there are some extremely short wavelength modes for which $y_{st}\gg1$. These modes will change the vacuum for $\tau\ll\tc$ so that the Bunch-Davies solution \eqref{zeta-SR-initial} gets some corrections. However, as we will show, only the modes close to the horizon scale will significantly feel the transition. Therefore, these extremely short wavelength modes are irrelevant for our purpose.

Using junction conditions \eqref{JC} to glue the two solutions \eqref{zeta-SR-initial} and \eqref{zeta-ST-final} at the conformal time $\tc$, we fix the coefficients~$c^{st}_\ell$ as
\begin{eqnarray}\label{cl-SR-ST}
c^{st}_\ell =  (-1)^{\ell} \frac{i\pi}{8} \frac{H_{sr}}{H_{st}} 
\bigg[ x_{\rm c} \Big(\frac{\beta_{sr}}{\beta_{st}}\Big)^{\frac{1}{2}}
{\rm H}_{\frac{1}{2}}^{(1)}\left(x_{\rm c}\right)
{\rm H}_{\frac{3}{4}}^{(\ell)}\left(y_{\rm c}\right)
- 2 y_{\rm c} 
\Big(\frac{\beta_{st}}{\beta_{sr}}\Big)^{\frac{1}{2}} \frac{H_{sr}}{H_{st}}
{\rm H}_{\frac{3}{2}}^{(1)}\left(x_{\rm c}\right)
{\rm H}_{-\frac{1}{4}}^{(\ell)}\left(y_{\rm c}\right)
\bigg] \,,
\end{eqnarray}
where $\cc$ denotes the value of the sound speed at the time of transition. Note also that we used $\nu=3/2=\mu$. Clearly, the amplitude for the power spectrum at $\tau\gg\tc$ will not be affected by this simplification. Moreover, we will see that the contribution to the spectral tilt from the slow-roll corrections in $\nu$ are suppressed as well.

For $\tau\gg\tc$, the mode function is given by Eq.~\eqref{zeta-ST-final} such that the coefficients $c^{st}_\ell$ are given by Eq.~\eqref{cl-SR-ST}. The corresponding dimensionless power spectrum is given by
\begin{equation}\label{PS-zeta}
\Delta^2_{\zeta}(k)|_{\tau\gg\tc}  = \frac{H_{st}^2}{2\pi^2\Mp^2} \frac{H_{st}^2k^3\tau^2}{\beta_{st}M^2}
\big{|} \bar{\zeta}^{st,f}_k \big{|}^2 \,.
\end{equation}
Substituting \eqref{zeta-ST-final} into the above relation, we find
\begin{equation}\label{PS-zeta-st}
\Delta^2_{\zeta}(k)|_{\tau\gg\tc} = \frac{H_{st}^2}{\Mp^2} \Big(\frac{H_{st}}{M}\Big)^{1/2}
\frac{1}{\beta_{st}{\tilde \alpha}_{st}^{3/2}}
\frac{\Gamma(\mu/2)^2}{\pi^3} | c^{st}_1 - c^{st}_2 |^2 y_{st}^{3/2-\mu} \, .
\end{equation}
Note that the coefficients $c^{st}_\ell$ given by \eqref{cl-SR-ST} depend on the scale through $x_{\rm c}$ and $y_{\rm c}$. Then, the spectral tilt at the stealth regime is given by
\begin{equation}\label{sr-tilt-f}
n_{s}^{sr,f} -1 =
3-2\mu + \frac{d\ln| c^{st}_1 - c^{st}_2 |^2}{d\ln{k}} \,.
\end{equation}

\subsection{From stealth to slow-roll}

Now we consider a transition from an initial stealth phase to a final slow-roll phase. 

For $\tau\ll\tc$, we have a stealth regime which corresponds to the ghost inflation scenario \cite{ArkaniHamed:2003uz}. The mode function is given by Eq.~\eqref{zeta-stealth} as
\begin{align}\label{zeta-ST-initial} 
\bar{\zeta}^{st,i}_k = 
\frac{\sqrt{-\pi\tau}}{2\sqrt{2}} {\rm H}^{(1)}_{\mu/2}\left(y_{st}\right) \,.
\end{align}
The corresponding dimensionless power spectrum, which is defined by Eq.~\eqref{PS-def}, turns out to be
\begin{equation}\label{PS-zeta-st-2}
\Delta^2_{\zeta}(k)|_{\tau\ll\tc} = \frac{H_{st}^2}{2\pi^2\Mp^2} 
\Big(\frac{H_{st}}{M}\Big)^{1/2} \frac{1}{\beta_{st}{\tilde\alpha}_{st}^{3/2}}
\Big( \frac{\Gamma(\mu/2)^2}{\pi} \Big) \Big( \frac{y_{st}}{2} \Big)^{3/2-\mu}
\, ,
\end{equation}
which leads to the following spectral tilt
\begin{equation}\label{st-tilt}
n_{s}^{st,i} -1 = 3-2\mu \,.
\end{equation}

For $\tau\gg\tc$, we have slow-roll inflation and the mode function takes the form
\begin{align}\label{zeta-SR-final} 
\bar{\zeta}^{sr,f}_k = 
\frac{\sqrt{- \pi \tau }}{2\sqrt{2}} \left[ 
c^{sr}_2 {\rm H}_{\nu}^{(1)}\left(x_{sr} \right)
+c^{sr}_1 {\rm H}_{\nu}^{(2)}\left(x_{sr} \right)\right] \,,
\end{align}
where coefficients $c^{sr}_\ell$ with $\ell=1,2$ are some constants. 
Applying the junction conditions \eqref{JC} we find
\begin{eqnarray}
c^{sr}_\ell = (-1)^{\ell} \frac{i\pi}{4} \frac{H_{st}}{H_{sr}} 
\bigg[ x_{\rm c} \Big(\frac{\beta_{sr}}{\beta_{st}}\Big)^{\frac{1}{2}} \frac{H_{st}}{H_{sr}}
{\rm H}_{\frac{3}{4}}^{(1)}\left(y_{\rm c}\right)
{\rm H}_{\frac{1}{2}}^{(\ell)}\left(x_{\rm c}\right)
- 2 y_{\rm c} 
\Big(\frac{\beta_{st}}{\beta_{sr}}\Big)^{\frac{1}{2}} 
{\rm H}_{-\frac{1}{4}}^{(1)}\left(y_{\rm c}\right)
{\rm H}_{\frac{3}{2}}^{(\ell)}\left(x_{\rm c}\right)
\bigg] \,,
\end{eqnarray}
where we have used $\nu=3/2=\mu$. The parameter $\vartheta:\mu \to \nu$ is plotted in the right panel of Fig.~\ref{Fig:vartheta-2}. 

Following the same steps as the previous subsection, we find the dimensionless power spectrum
\begin{equation}\label{PS-zeta-sr-2}
\Delta^2_{\zeta}(k)|_{\tau\gg\tc} = 
\frac{H_{sr}^2}{8\pi^2 \Mp^2} \frac{H_{sr}^2}{M^2} 
\frac{1}{\beta_{sr}\csr^3} \Big( \frac{\Gamma(\nu)}{\Gamma(3/2)}\Big)^2 | c^{sr}_1 - c^{sr}_2 |^2 x_{sr}^{3-2\nu} \, .
\end{equation}
The spectral tilt in this case is given by
\begin{equation}\label{st-tilt-f}
n_{s}^{st,f} -1 = 
3-2\nu + \frac{d\ln| c^{sr}_1 - c^{sr}_2 |^2}{d\ln{k}} \,.
\end{equation}

\section{Possibility of formation of PBHs}
\label{sec:PBHs}

Until now, we did not consider any applications of our setup. We only formulated a general single field inflationary scenario with $0\leq\cs\leq1$ which allows for transitions $\cs:\csr\to0$ and $\cs:0\to\csr$. In this section, we look for the possibility of the formation of primordial black holes (PBHs) on small scales. In order to form PBHs, we need to enhance the power spectrum sufficiently on sub-CMB scales. In the canonical inflation, this requirement leads to a small value of $\epsilon$ or an ${\cal O}(1)$ violation of slow-roll~\cite{Motohashi:2017kbs,Passaglia:2018ixg}. In the case of k-inflation, instead of requiring small $\epsilon$, one may require small $\cs$ to enhance the power spectrum~\cite{Kamenshchik:2018sig,Kamenshchik:2021kcw}. However, as we mentioned, with small $\cs$, the higher derivative scordatura term becomes important to avoid the strong coupling issue. Thus, one may need to take into account the scordatura corrections since the estimation without the scordatura would be changed significantly. Below we shall clarify this issue.

Looking at the dispersion relation \eqref{DR-total}, we find that the natural choice is the case of transition from the slow-roll to the stealth when the $k^4$ term dominates on small scales. Indeed, this possibility is studied very recently in the context of EFT of inflation \cite{Ballesteros:2018wlw,Ashoorioon:2019xqc,Ballesteros:2021fsp}. Here, however, we have an explicit model in hand which we have presented in subsection \ref{subsec-SRtoST}.

Suppose that modes corresponding to the CMB scales leave the horizon before some conformal time $\tcmb$. The transition from slow-roll regime to the stealth regime can take place at any time during inflation so that we can consider both cases of $\tc<\tcmb$ and $\tc>\tcmb$, where $\tc$ is the conformal time for the transition. In order to allow for sufficiently large enhancement of the power spectrum on small scales, we consider the case $\tc>\tcmb$. In this case, we do not see any observable feature of the transition in the CMB spectrum. We then look for the effects of the transition on the sub-CMB modes which are subhorizon by the time $\tcmb$. If we assume ${\cal N}_{\rm c}-{\cal N}_{\ast}={\cal O}(10)$ where ${\cal N}_{\rm c}=-\ln(-H_{\rm c}\tc)$ and ${\cal N}_{\ast}=-\ln(-H_{\ast}\tcmb)$ are the number of e-folds at the times ${\tau}_{\rm c}$ and $\tcmb$ respectively, we will have $\tau_{\rm {c}}\gg\tcmb$ . In this regard, for the CMB scales, the mode function is given by the slow-roll one, i.e., Eq.~\eqref{zeta-SR-initial}.

In order to find the amplitude of the power spectrum at $\tau\gg\tc$, we need to look at sub-CMB modes. The modes which leave the sound horizon before the time $\tcmb$ satisfy $-\csr k\tcmb\ll1$. Therefore we define the CMB scale as
\begin{eqnarray}
k_{\ast,sr} \equiv \frac{1}{-\csr\tcmb} \,.
\end{eqnarray}
The modes $k<k_{\ast,sr}$ become superhorizon before $\tcmb$ while the modes $k>k_{\ast,sr}$ become superhorizon after $\tcmb$. Let us also define another scale
\begin{eqnarray}
k_{{\rm c},sr} \equiv \frac{1}{-\cc\tc} \,.
\end{eqnarray}
From the above definitions, we find that $x_{\rm c}\leq1$ for the modes $k_{\ast,sr}\leq k \leq k_{{\rm c},sr}$ while $x_{\rm c}>1$ for the modes $k>k_{{\rm c},sr}$. Note that $k_{{\rm c},sr}/k_{\ast,sr}=(\csr/\cc)(|\tcmb|/|\tc|)\gg1$ and the window $k_{\ast,sr}\leq k \leq k_{{\rm c},sr}$ is large.

Having considered the case $\tc\gg\tcmb$, $H_{sr}$ will be fixed by the COBE normalization. The slow-roll parameters $\epsilon_{sr}$, $\eta_{sr}$, and $s_{sr}$ will be also fixed by CMB observations with the typical value of ${\cal O}(10^{-2})$. At the time $\tau\gg\tc$, $y_{\rm c}\ll1$ for the modes $k_{\ast,sr}\leq k \leq k_{{\rm c},sr}$ with $x_{\rm c}\leq1$. Approximating the Hankel functions in Eq.~\eqref{cl-SR-ST} for $y_{\rm c}\ll1$ we find 
\begin{eqnarray}\label{cl-SR-to-ST}
| c^{st}_1 - c^{st}_2 |^2 \approx 
\frac{9\pi}{64\Gamma(7/4)^2} \frac{\beta_{st}}{\beta_{sr}} \Big(\frac{H_{sr}}{H_{st}}\Big)^{4} 
\Big( \frac{{\tilde \alpha}_{\rm c}H_{\rm c}}{M} \Big)^{3/2} \frac{1}{\cc^3}
\Big[ \Big( 1 - \frac{\beta_{sr}}{3\beta_{st}} \frac{H_{st}}{H_{sr}} x_{\rm c}^2 \Big)^2 
+ x_{\rm c}^2 \Big] \,.
\end{eqnarray}
In obtaining the above result, we substituted $\mu=3/2$. Since 
\begin{eqnarray}\label{yc}
y_{\rm c} = ({\tilde\alpha}_{\rm c} H_{\rm c}/2M\cc^2) x_{\rm c}^2 \,,
\end{eqnarray}
we can have $x_{\rm c}\gg1$ while $y_{\rm c}\ll1$ for some modes $k>k_{{\rm c},sr}$. Thus, the above results are valid for those modes as well. In order to be more precise, we define another scale
\begin{eqnarray}
k_{{\rm c},st} \equiv \frac{1}{-\gamma_{\rm c}\tc} \,; 
\hspace{1cm}
\gamma \equiv \Big( \frac{{\tilde\alpha}H}{2M} \Big)^{1/2} \,.
\end{eqnarray}
From the above definition, we find $y_{\rm c}\leq1$ for the modes $k \leq k_{{\rm c},st}$ while $y_{\rm c}>1$ for the modes $k>k_{{\rm c},st}$. Therefore, $x_{\rm c}\gg1$ while $y_{\rm c}\ll1$ for the modes $k_{{\rm c},sr}<k\ll {k}_{{\rm c},st}$. Note that the ratio ${k}_{{\rm c},st}/{k}_{{\rm c},sr}=\cc(2M/{\tilde\alpha_{\rm c}H_{\rm c}})^{1/2}$ is large since $M/H_{\rm c}\gg1$.

For the modes $k_{\ast,sr}\leq k \leq k_{{\rm c},sr}$, we have $x_{\rm c}\leq1$ and with good accuracy, the bracket in Eq.~\eqref{cl-SR-to-ST} becomes of the order of unity and therefore it is not possible to enhance the power spectrum. On the other hand, for the modes $k_{{\rm c},sr}<k\ll {k}_{{\rm c},st}$ with $x_{\rm c}\gg1$ but still $y_{\rm c}\ll1$, the term quartic in $x_{\rm c}$ in the bracket in Eq.~\eqref{cl-SR-to-ST} dominates and Eq.~\eqref{cl-SR-ST} gives
\begin{align}\label{cl-SR-to-ST-x-large}
&| c^{st}_1 - c^{st}_2 |^2 \approx 
\frac{\pi}{64\Gamma(7/4)^2} \frac{\beta_{st}}{\beta_{sr}} 
\Big( \frac{{\tilde \alpha}_{\rm c}H_{\rm c}}{M} \Big)^{3/2} \frac{1}{\cc^3}
\Big(\frac{H_{sr}}{H_{st}}\Big)^{2} x_{\rm c}^4 \,;
&\mbox{for} \hspace{.5cm} k_{{\rm c},sr}<k\ll {k}_{{\rm c},st} \,.
\end{align}
Substituting the above result in \eqref{PS-zeta-st}, we find
\begin{equation}\label{PS-zeta-st-sim}
\Delta^2_{\zeta}(k)|_{\tau\gg\tc} = \frac{H_{st}^2}{36 \pi^2\Mp^2} \Big(\frac{H_{sr}}{M}\Big)^{2}
\frac{1}{\beta_{sr}\cc^3} 
\Big( \frac{{\tilde \alpha}_{\rm c}H_{\rm c}}{{\tilde \alpha}_{st} H_{st} } \Big)^{3/2}
x_{\rm c}^4 \, ,
\end{equation}
where we have substituted $\mu=3/2=\nu$.  From \eqref{PS-zeta-st-sim} and \eqref{PS-zeta-sr} we then find
\begin{align}\label{PS-ratio-SR-to-ST-2}
&\frac{\Delta^2_{\zeta}(k)|_{\tau\gg\tc}}{\Delta^2_{\zeta}(k)|_{\tau\ll\tc} } \approx 
\frac{1}{9} \Big( \frac{{\tilde \alpha}_{\rm c}H_{\rm c}}{{\tilde \alpha}_{st} H_{st} } \Big)^{3/2} 
\Big( \frac{H_{st}}{H_{sr}} \Big)^{2} \frac{\csr^3} {\cc^{3}}
x_{\rm c}^4 \,;
&\mbox{for} \hspace{.5cm} k_{{\rm c},sr}<k\ll {k}_{{\rm c},st} \,.
\end{align} 

Let us now look at the spectral tilt of the power spectrum after the time $\tau\gg\tc$. The power-law term $y_{st}^{3/2-\mu}$ in \eqref{PS-zeta-st} gives the usual almost scale-invariant spectral tilt. However, the power spectrum depends on the scale through $| c^{st}_1 - c^{st}_2 |^2$ as well. The corresponding contribution to the spectral tilt is
\begin{align}\label{tilt-SR-to-ST}
&n^{st,f}_s -1 \approx \frac{d\ln| c^{st}_1 - c^{st}_2 |^2}{d\ln{k}} = 4 \,,
&&\mbox{for} \hspace{.5cm} k_{{\rm c},sr}<k\ll {k}_{{\rm c},st} ,
\end{align}
where in the last step we have considered \eqref{cl-SR-to-ST-x-large} for the short wavelength modes $k_{{\rm c},sr}<k\ll {k}_{{\rm c},st}$. Therefore, the power spectrum for the curvature perturbations is blue-tilted on small scales and it is possible to enhance it for the modes $k_{{\rm c},sr}<k\ll {k}_{{\rm c},st}$. 

Now, the question is whether the enhancement is large enough to allow for the formation of PBHs. In order to answer this question, we note that the spectrum is blue-tilted and, therefore, it has a peak for the maximum value of $x_{\rm c}$. From Eq.~\eqref{yc}, we find that $x_{\rm c} \ll (2M/{\tilde \alpha}_{\rm c} H_{\rm c})^{1/2} \cc$ for $y_{\rm c}\ll1$. Thus, in order to find an upper bound for the power spectrum, we evaluate it for the maximum value $x_{\rm c} = (2M/{\tilde \alpha}_{\rm c} H_{\rm c})^{1/2} \cc$. Although this is not possible within our regime of approximation, it is enough to prove that the power spectrum cannot be enhanced even for this large value. Substituting this value in Eq.~\eqref{PS-ratio-SR-to-ST-2}, we find
\begin{align}\label{PS-ratio-SR-to-ST-max}
&\frac{\Delta^2_{\zeta}(k)|_{\tau\gg\tc}}{\Delta^2_{\zeta}(k)|_{\tau\ll\tc} } \approx 
\frac{4}{9} \Big( \frac{H_{st}}{H_{sr}} \Big)^{2} \frac{\csr^3 \cc}{\sqrt{1-\cc^2}}
\beta_{st} \Big( \frac{\beta_{\rm c}}{\beta_{st}} \Big)^{1/4}
\Big( \frac{M}{H_{st}} \Big)^2 \frac{1}{\alpha} \,;
&\mbox{for} \hspace{.5cm} x_{\rm c} 
= \Big(\frac{2M}{{\tilde \alpha}_{\rm c} H_{\rm c}}\Big)^{1/2} \cc \,,
\end{align} 
where we have used the fact that ${\tilde \alpha}_{\rm c}/{\tilde\alpha}_{st}=\sqrt{\beta_{st}/\beta_{\rm c}}(1-\cc^2)$ and we have also substituted \eqref{alpha-tilde-stealth}. The enhancement can happen due to the large values of the last part $(M/H_{st})^2 \alpha^{-1}$. For the typical values of $H_{st}$ and $M$, $(M/H_{st})^2\lesssim{\cal O}(10^4)$. 
In order to allow for the PBHs to be dark matter, the power spectrum needs to be enhanced up to $\Delta^2_{\zeta}(k)|_{\tau\gg\tc}={\cal O}(10^{-2})$~\cite{Motohashi:2017kbs}. 
Therefore, to achieve $\Delta^2_{\zeta}(k)|_{\tau\gg\tc}/\Delta^2_{\zeta}(k)|_{\tau\ll\tc} = {\cal O}(10^{7})$,
we have to choose $\alpha\lesssim {\cal O}(10^{-3})$.
Thus, even for the large value of $x_{\rm c} = (2M/{\tilde \alpha}_{\rm c} H_{\rm c})^{1/2} \cc$, we need to fine-tune the scordatura coupling constant $\alpha$. More precise numerical estimation shows that for the typical values of $x_{\rm c}\ll (2M/{\tilde \alpha}_{\rm c} H_{\rm c})^{1/2} \cc$, we have to assume even $\alpha\lesssim {\cal O}(10^{-7})$. Since the EFT cutoff scale is $M/\sqrt{\alpha}$, small $\alpha$ makes the EFT cutoff scale high, and eventually contradicts with $M/M_{\rm Pl}\ll 1$.  
In other words, our result suggests that large enhancement of the power spectrum is likely to be prevented by the contribution from the scordatura term. 
Therefore, the analysis of PBH formation relying on $\cs \ll 1$, for which the scordatura term becomes nonnegligible, needs to be revisited to take into account the scordatura corrections.

Finally, let us make a comment on the extremely short wavelength modes. One may roughly think that these modes will give the dominant contribution as the $k^4$ term dominates $k^2$ term in the dispersion relation \eqref{DR-total} for the extremely short wavelength modes. However, this is not the case. First of all, as we mentioned above, the Bunch-Davies solution \eqref{zeta-SR-initial} would be affected by these modes even for $\tau\ll\tc$ (see for instance Ref. \cite{Ashoorioon:2017toq,Ashoorioon:2018ocr}). Thus, our setup would be changed accordingly if we want to study their behavior in a precise manner. Actually, we do not need to do that. The reason is that, if we consider the modes for which not only $x_{\rm c}\gg1$ but also $y_{\rm c}\gg1$, we can directly confirm that the spectral tilt becomes smaller for large momenta. In this regard, the power spectrum becomes blue-tilted only for the modes $k_{{\rm c},sr}<k\ll {k}_{{\rm c},st}$. Note that neither $y_{\rm c} = 1$ nor $x_{\rm c} = 1$ define horizon exit as the setup is neither stealth nor slow-roll at the time of transition. This result can be intuitively understood if we note that the scales $k_{{\rm c},sr}<k\ll {k}_{{\rm c},st}$ are indeed around the horizon scale. Thus, those modes that are near the horizon scale experience the transition more than other smaller or larger modes. 

\section{Summary}
\label{sec:sum}

Inflationary scenarios based on the k-inflation models have interesting observational features in the regime of small sound speed $\cs\ll1$ for the curvature perturbations. 
Since there is a lower bound on the sound speed below which the linear perturbation theory based on the k-inflation action is no longer applicable at the horizon scale, one needs to include a higher dimensional operator around the background with $\cs \simeq 0$ as originally introduced in ghost condensation/inflation. Roles of the higher dimensional operator around backgrounds with $\cs \simeq 0$ were recently revisited under the name of ``scordatura''. 

In this paper, we have studied a self-consistent inflationary model that exhibits a transition between a slow-roll k-inflation with $\cs$ of order unity and a ghost inflation with $\cs \simeq 0$. 
We have constructed a novel inflationary model for the transition between the two phases, which are unified smoothly by appropriately taking into account the higher derivative scordatura term. In this setup, the parameter space of the k-inflation is enlarged and one can achieve the whole range of $0\leq\cs\leq1$ avoiding strong coupling and gradient instability by virtue of the scordatura mechanism. We have first clarified a background configuration which supports a transition between a slow-roll k-inflation and a ghost inflation. Then, using the background dynamics, we have obtained the power spectrum of the curvature perturbations at the end of inflation. 

These results can be used for different phenomenological purposes. As an application, we have explored the possibility of the formation of PBHs as dark matter.  
We have focused on the case of a transition from a slow-roll k-inflation to a ghost inflation and assumed that a transition takes place after the horizon exit of the CMB scales. In this regard, on the CMB scales the model prediction is the same as the standard slow-roll k-inflation, while we can address the enhancement of the power spectrum on sub-CMB scales by the ghost inflation. As a result, we have found that, while it is possible to enhance the power spectrum on sub-CMB scales, the enhancement is not sufficiently large to allow for the formation of PBHs as the origin of all dark matter in the universe. 
Our results are robust as far as the low energy EFT describing perturbations is required to be weakly coupled all the way down to the $\cs \simeq 0$ regime, as known from the theories of ghost condensation/inflation~\cite{ArkaniHamed:2003uy,ArkaniHamed:2003uz} and scordatura~\cite{Motohashi:2019ymr,Gorji:2020bfl}. This makes it challenging to build a self-consistent inflationary model for producing the PBHs as the origin of all dark matter in the universe solely from small sound speed.

\vspace{.5cm}
{\bf Acknowledgments:} The work of M.A.G.\ was supported by Japan Society for the Promotion of Science (JSPS) Grants-in-Aid for international research fellow No.\ 19F19313.
H.M.\ was supported by JSPS Grant-in-Aid for Scientific Research (KAKENHI) Grant No.\ JP18K13565. 
S.M.'s work was supported in part by Japan Society for the Promotion of Science Grants-in-Aid for Scientific Research No.\ 17H02890, No.\ 17H06359, and by World Premier International Research Center Initiative, MEXT, Japan. 

\vspace{0.2cm}

\bibliographystyle{JHEPmod}
\bibliography{ref}

\end{document}